\documentclass[journal]{new-aiaa}
\usepackage{authblk}
\usepackage{graphicx} 
\usepackage{amsmath}
\usepackage{bm}
\usepackage{flafter}
\usepackage[title]{appendix}
\usepackage{xcolor}
\usepackage{algorithm}
\usepackage{algpseudocode}

\usepackage{subfigure}

\newcommand{\EE}[1]{\mathbb{E}\left[#1\right]}
\newcommand{\Var}[1]{\mathbb{V}ar\!\left[#1\right]}

\newcommand{\avec}{\underline{\alpha}}
\newcommand{\zvec}{\underline{\mathbf{z}}}

\newcommand{\mcest}{\hat{Q}}

\newcommand{\acvest}{\tilde{Q}}

\newcommand{\diffvec}{\underline{\Delta}}

\newcommand{\covmat}{\mathbf{C}}
\newcommand{\covvec}{\mathbf{c}}

\newcommand{\Fmat}{\mathbf{F}}
\newcommand{\fvec}{\mathbf{f}}

\newcommand{\sset}[2]{\mathbf{z}_{#1}^{#2}}

\newcommand{\covop}[2]{\mathbb{C}ov\left[#1,#2\right]}
\newcommand{\corrop}[2]{\mathbb{C}orr\left[#1,#2\right]}

\DeclareMathOperator*{\argmin}{arg\,min}

\newcommand{\hyper}{\underline{\beta}}  
\newcommand{\Pmat}{\mathbf{P}}
\newcommand{\rhovec}{\underline{\rho}}

\title{Automated Model Tuning for Multifidelity Uncertainty Propagation in Trajectory Simulation}
\author{James E. Warner \footnote[1]{Research Computer Scientist, Durability, Damage Tolerance, and Reliability Branch, M/S 188E; Corresponding author (james.e.warner@nasa.gov).} and Geoffrey F. Bomarito \footnote[2]{Materials Engineer, Durability, Damage Tolerance, and Reliability Branch, M/S 188E.}}
\affil{NASA Langley Research Center, Hampton, VA 23681, USA}
\author{Gianluca Geraci\footnote[3]{Principal Member of Technical Staff, Optimization and 
		Uncertainty Quantification, Sandia National Laboratories, MS1318, Albuquerque, NM 87185, AIAA Senior Member.} and Michael S. Eldred\footnote[4]{Distinguished Member of Technical Staff,
    Optimization and Uncertainty Quantification, Sandia National
    Laboratories, MS1318, Albuquerque, NM 87185,
    AIAA Associate Fellow.}}
\affil{Sandia National Laboratories, Albuquerque, NM 87185, USA}   

\date{}

\begin{document}

\maketitle

\begin{abstract}

Multifidelity uncertainty propagation combines the efficiency of low-fidelity models with the accuracy of a high-fidelity model to construct statistical estimators of quantities of interest. It is well known that the effectiveness of such methods depends crucially on the relative correlations and computational costs of the available computational models. However, the question of how to automatically tune low-fidelity models to maximize performance remains an open area of research. This work investigates automated model tuning, which optimizes model hyperparameters to minimize estimator variance within a target computational budget.  Focusing on multifidelity trajectory simulation estimators, the cost-versus-precision tradeoff enabled by this approach is demonstrated in a practical, online setting where upfront tuning costs cannot be amortized. Using a real-world entry, descent, and landing example, it is shown that automated model tuning largely outperforms  hand-tuned models even when the overall computational budget is relatively low.  Furthermore, for scenarios where the computational budget is large, model tuning solutions can approach the best-case multifidelity estimator performance where optimal model hyperparameters are known \textit{a priori}. Recommendations for applying model tuning in practice are provided and avenues for enabling adoption of such approaches for budget-constrained problems are highlighted.

\end{abstract}

\section{Introduction}

Trajectory simulation for entry, descent, and landing (EDL) is a cornerstone capability for the National Aeronautics and Space Administration's (NASA's) design of space missions. Accurately and precisely predicting the characteristics and risks associated with a trajectory thus has a huge significance for the agency. For example, NASA aims to deliver several payloads to Mars within a target landing radius of 50 meters \cite{toups2015pioneering} in order to safely coordinate multiple landings in support of a single mission. This aggressive target is currently orders of magnitude more precise than the current state-of-the-art \cite{dwyer2017entry}.  Achieving such precision is complicated by numerous sources of uncertainty, including atmospheric variability, navigation errors, and modeling approximations, all of which contribute to landing dispersions when propagated through trajectory simulation models. Consequently, advancements in uncertainty propagation methods is one key avenue for improving EDL performance in support of future missions.

Multifidelity uncertainty propagation has been shown to enable more precise and efficient estimates of a trajectory's quantities of interest (QoIs) \cite{warner_multi-model_2021} when compared to the standard Monte Carlo (MC) approach.  Multifidelity uncertainty propagation techniques leverage low-fidelity models (fast but biased) to deliver efficient statistical estimates of a high-fidelity model's (slow but accurate) QoIs. Multifidelity uncertainty propagation started, in the sampling context, with Heinrich~\cite{Heinrich2001} and Giles~\cite{Giles2008} who pioneered the use of coarser approximations for the solution of stochastic integral equations and stochastic differential equations, respectively. Building upon these seminal contributions, a large multi-level MC (MLMC) literature developed, see Giles' review paper~\cite{giles2015multilevel}. Several other related approaches were later proposed over the last decade, including multi-index MC~\cite{HajiAli2015,HajiAli2016}, multifidelity MC (MFMC)~\cite{Ng2014,peherstorfer2016optimal,Peherstorfer2018}, multi-level control variate MC approaches~\cite{Nobile2015,Fairbanks2017,Geraci2017} and, most recently, Approximate Control Variate (ACV)~\cite{gorodetsky2020generalized,BOMARITO2022110882,gorodetsky2024grouped} approaches and multi-level best linear unbiased estimators (MLBLUE)~\cite{schaden2020multilevel,Schaden2021}. Despite differences in the construction of the classes of multifidelity estimators described above, virtually all of them rely on low-fidelity models directly derived from the high-fidelity model. For example, it is customary to use less accurate time integration, grid coarsening, and/or introduction of simplified physics.  As such, low-fidelity models typically have hyperparameters that can affect both their accuracy and computational cost.  The selection of these hyperparameters directly impacts the efficiency and precision of multifidelity estimators, making low-fidelity model tuning (i.e., the automated selection of optimal hyperparameters) an important avenue of research\footnote{An alternative (and possibly complementary) approach to achieving similar goals is to rely on the construction of shared low-dimensional spaces to augment the correlation among low- and high-fidelity models through a mapping of the input spaces, see ~\cite{Geraci2018,Geraci2018b,Zeng2023,Zeng2025,zanoni2024,Zanoni2025} for more details.}.

Automated model tuning has been recently demonstrated in the context of trajectory simulation \cite{bomarito_improving_2022,thompson_strategies_2023}, thermal battery analysis \cite{adams2022deployment}, and high-speed flows~\cite{Gruber2026} but the area remains largely unexplored beyond these studies. \citet{bomarito_improving_2022} first posed model tuning as an optimization problem where model hyperparameters that minimize multifidelity estimator variance are sought. However, this work mainly focused on exploring the potential increase in speed and precision for multifidelity trajectory simulation rather than providing a practical implementation of the approach.  \citet{thompson_strategies_2023} and \citet{adams2022deployment} later employed Efficient Global Optimization (EGO) as a robust solution strategy, but both studies considered the model tuning optimization problem in isolation rather than analyzing its impact on cost and precision within a budget-constrained multifidelity uncertainty propagation analysis. In particular, multifidelity estimates of statistics were not actually formed using the tuned low-fidelity models; instead the performance was gauged through the optimal solutions for \textit{projected} estimator variance. Furthermore, the trajectory simulation study of \citet{thompson_strategies_2023} was limited to single tunable hyperparameter and only considered one QoI. Thus, a comprehensive study of the cost-versus-precision tradeoff provided by allocating computational budget to automated model tuning is lacking and its potential for improving multifidelity trajectory simulation is not fully understood.

This paper provides an in-depth study of automated model tuning for multifidelity uncertainty propagation, representing the culmination of a line of previous work on the topic \cite{bomarito_improving_2022,thompson_strategies_2023}. A real-world entry, descent, and landing example - the Adaptable, Deployable Entry and Placement Technology (ADEPT) project test flight - \cite{cassell2018adept,dutta2019flight} is considered to develop practical recommendations for the approach.  This case study uses the Program to Optimize Simulated Trajectories II (POST2), a legacy NASA Langley Research Center code, as a variable-fidelity simulator \cite{lugo2017launch}.  The goal here is to estimate QoIs such as landing location and touchdown time given a set of uncertainties associated with initial vehicle state, atmospheric conditions, etc., under a limited computational budget where upfront tuning costs cannot be ignored. After optimal hyperparameters are identified with model tuning, QoI estimators are formed using an iterated ACV approach \cite{adams2022deployment} that adaptively estimates model QoI statistics (correlations, variances) that are required for multifidelity estimation. To explore the cost-versus-precision tradeoff associated with model tuning, results are benchmarked against baseline ACV estimators that use hand-tuned models  \cite{warner_multi-model_2021} as well as the best-case ACV performance where optimal model hyperparameters and QoI statistics are known \textit{a priori}. Sandia National Laboratories' flagship uncertainty quantification code, Dakota  \cite{dakota2024}, is used as the basis for the implementation of both ACV estimators and model tuning optimization. The primary contributions of this work are summarized as follows:
\begin{itemize}
	\item A rigorous study of the cost-versus-precision tradeoff from allocating upfront computation to automated model tuning by comparing the approach to both baseline and best-case performance within an end-to-end uncertainty propagation analysis for varying computational budgets. 
	\item An improvement to the state of the art for multifidelity trajectory simulation estimators demonstrated through comparisons to original work in this area \cite{warner_multi-model_2021}.
	\item The first demonstration of an \textit{online} multifidelity uncertainty propagation procedure where neither optimal low-fidelity model hyperparameters nor knowledge of model statistics are known \textit{a priori} using automated model tuning and iterated ACV.
	\item Recommendations for applying automated model tuning to practical problems and avenues for future work to improve performance.
\end{itemize}

\section{Multifidelity Estimators and Model Tuning}
\label{sec: model_tuning_problem}

In this section, the Approximate Control Variate (ACV) multifidelity framework is briefly introduced as based on previous work~\cite{gorodetsky2020generalized,BOMARITO2022110882}. Next, the so-called \emph{model tuning} problem, which constitutes the main contribution of the paper, is introduced and discussed. It is important to note that the model tuning idea can be adopted with minor modification in virtually any existing multifidelity sampling estimator, e.g., MFMC~\cite{peherstorfer2016optimal} or MLBLUE~\cite{schaden2020multilevel,Schaden2021}, but the ACV framework is emphasized here due to its flexibility. ACV has unified existing methods, MLMC~\cite{giles2015multilevel} and MFMC, it encompasses a large group of possible estimators \cite{BOMARITO2022110882,gorodetsky2024grouped}, and it also has direct relationships with MLBLUE~\cite{schaden2020multilevel,Schaden2021,gorodetsky2024grouped}. The idea of all multifidelity estimators is to optimally combine realizations from an ensemble of models with varying 
cost and accuracy to increase the statistical precision of targeted quantities for the high-fidelity model, for which only a limited number of simulations can be acquired. In this presentation, the focus is on the estimation of an expected value, but the extension to higher order moments or densities is also possible~\cite{AGuilmard2023,Menhorn2024,Dixon2026}. 

Let $Q: \mathbb{R}^d \rightarrow \mathbb{R}$ represent the mapping from a vector of input $Z \in \mathbb{R}^d$ to a scalar quantity of interest (QoI) for a high-fidelity model, while $Q_i$ represent $M$ suitable low-fidelity approximations of $Q$. An Optimal Control Variate (OCV) estimator ~\cite{Lavenberg1978,Lavenberg1981,Lavenberg1982}, is a Monte Carlo-based linear control variate estimator for the expected value of $Q$, hereafter denoted as $\EE{Q}$, with respect to $Z$.  It can be written as
\begin{equation} \label{eq:OCV_est}
\hat{Q}^{OCV} = \mcest(\sset{}{}) + \sum_{i=1}^M \alpha_i \left( \mcest_i(\sset{}{}) - \mu_i \right),
\end{equation}
where the symbol $\hat{\cdot}$ denotes a MC sampling estimator, e.g.,
\begin{equation}\label{eq: MC_estimator}
\mcest(\sset{}{}) = \frac{1}{N} \sum_{z \in \sset{}{}} Q(z)
\end{equation}
which is based on a set, $\sset{}{}$, of $N$ independent realizations of $Z$ (identical definitions apply for $Q_i$ with $i=1,\dots,M$), $\alpha_i$ represent weights to be optimized, and $\mu_i$ are the known expected values for the $M$ low-fidelity models, i.e., $\EE{Q_i} = \mu_i$. Sampling estimators are random variables because they are based on finite realizations of the input, but it is easy to show that they are unbiased, i.e., $\EE{\hat{Q}^{OCV}}=\EE{\mcest(\sset{}{})}=\EE{Q}$ for any choice of $\alpha_i \in \mathbb{R}$. On the other hand, the OCV variance can be minimized when the weights are chosen as $\avec = \left[ \alpha_1, \dots, \alpha_M \right]^\mathrm{T} = - \covmat^{-1} \covvec$, where $\covmat \in \mathbb{R}^{M \times M}$ denotes the covariance matrix among $Q_i$ and $\covvec \in \mathbb{R}^M$ denotes the covariance vector between $Q$ and $Q_i$. This variance can be expressed as 
\begin{equation} \label{eq:OCV_var}
\Var{\hat{Q}^{OCV}} = \Var{\mcest(\sset{}{})} \left( 1 - \frac{\covvec^\mathrm{T} \covmat^{-1} \covvec}{\Var{Q}} \right) = \Var{\mcest(\sset{}{})} \left( 1 - R^2_{OCV} \right),
\end{equation}
where $\Var{\mcest(\sset{}{})} = \dfrac{\Var{Q}}{N}$ is the MC estimator variance and $0 \leq R^2_{OCV} \leq 1$ is a measure of the variance reduction with respect to the MC estimator's variance. The OCV framework demonstrates that, by providing correlated low-fidelity models, it is possible in principle to reduce the MC variance; however, for realistic problems, the expected values $\mu_i$ need to be estimated and a sampling strategy needs to be chosen. 

The ACV framework encompasses several estimators and is defined by using $N_i = r_i N$ samples for each low-fidelity model and by partitioning this set in two ordered subsets $\sset{i}{1}$ and $\sset{i}{2}$ so that $\sset{i}{} = \sset{i}{1} \cup \sset{i}{2}$. The subsets are not required to be disjoint, i.e., it is admissible to have $\sset{i}{1} \cap \sset{i}{2} \neq 0$. 
The generic ACV estimator can then be written as
\begin{equation}\label{eq: approximate_CV}
\begin{split}
 \acvest\left(\avec,\zvec\right) &= \mcest(\sset{}{}) + \sum_{i=1}^M \alpha_i \left( \mcest_i(\sset{i}{1}) - \mcest_i(\sset{i}{2}) \right) \\
                                 &= \mcest(\sset{}{}) + \avec^\mathrm{T} \diffvec,
\end{split}
\end{equation}
where the vector of difference estimators is defined as $\diffvec = \left[ \Delta_1=\mcest_i(\sset{1}{1}) - \mcest_i(\sset{1}{2}), \dots,  \Delta_M=\mcest_i(\sset{M}{1}) - \mcest_i(\sset{M}{2}) \right]^\mathrm{T}$ and each term is obtained as the difference of two MC estimators for the $i$th model based on $\sset{i}{1}$ and $\sset{i}{2}$. Several estimators can be obtained by prescribing pre-determined relationships between $\zvec$, $\sset{i}{1}$, and $\sset{i}{2}$; MLMC, MFMC, and two ACV variants, namely ACV-IS and ACV-MF, are defined in Table~\ref{tab: summary_estimators_ACV} (from~\cite{gorodetsky2020generalized}). It is important to note that MLMC is recovered by introducing the additional simplification of fixing the weights $\avec=-1$, whereas a weighted MLMC estimator can be obtained if these weights are optimized while still keeping the same MLMC partitioning for $\sset{i}{1}$, and $\sset{i}{2}$, as discussed in~\cite{gorodetsky2020generalized}.

\renewcommand{\arraystretch}{1.1}
\begin{table}[h!]
  \caption{Sampling schemes for ACV estimators from~\cite{gorodetsky2020generalized}.}
  \centering
  \begin{tabular}{cccc}
    \hline \hline
    Algorithm & Relation between $\sset{}{}$ and $\sset{i}{}$ &$\sset{i}{1}$ & $\sset{i}{2}$  \\
    \hline
    MLMC~\cite{giles2015multilevel}         & $\sset{1}{1} = \sset{}{}$, $\sset{}{} \cap \sset{i}{} = \emptyset$ for $i>1$ & $\sset{i-1}{2}$  & $\sset{i}{} \setminus \sset{i}{1}$   \\ 
    MFMC~\cite{peherstorfer2016optimal} & $\sset{i}{} \supset \sset{}{}$ for all $i$ & $\sset{i-1}{}$    & $\sset{i}{}$   \\
    ACV-IS   & $\sset{}{} \cap \sset{i}{} = \sset{i}{1}$, $\sset{i}{2} \cap \sset{j}{2} = \emptyset$ for $1\leq i \neq j$& $\sset{}{}$ & $ \sset{i}{} \setminus \sset{i}{1} $ \\ 
    ACV-MF   & $\sset{}{} \cap (\sset{i}{}\setminus\sset{i}{1}) = \emptyset$ & $\sset{}{}$ &  $\sset{i}{}$  \\ 
    \hline \hline
  \end{tabular}
  \label{tab: summary_estimators_ACV}
\end{table}

It can be shown that the optimal values for control variate weights are~\cite{gorodetsky2020generalized}
\begin{equation}\label{eq: optimal_alpha}
\avec^{opt} = - \covop{\diffvec(\zvec)}{\diffvec(\zvec)}^{-1} \covop{\diffvec(\zvec)}{\mcest(\zvec)},
\end{equation}
while the corresponding ACV variance is given by \cite{thompson_strategies_2023}
\begin{equation}\label{eq: acv_variance}
\begin{split}
 \Var{\acvest\left(\zvec \right)}  &= \Var{\mcest(\zvec)}  - 
                                     \covop{\diffvec(\zvec)}{\mcest(\zvec)}^\mathrm{T} \covop{\diffvec(\zvec)}{\diffvec(\zvec)}^{-1} 
                                     \covop{\diffvec(\zvec)}{\mcest(\zvec)} \\
 &= \frac{\Var{Q}}{N} \left( 1 - \left[\fvec(\zvec) \circ \rhovec \right]^\mathrm{T}\left[\Fmat(\zvec) \circ \Pmat \right]^{-1}\left[\fvec(\zvec) \circ \rhovec \right] \right) \\
 &= \frac{\Var{Q}}{N} \left( 1 - R^2_{ACV} \right)
\end{split}
\end{equation}
where $\Pmat$ is the matrix of the correlations among low-fidelity models (i.e., $P_{ij} = \corrop{Q_i}{Q_j}$), the $\rhovec$ vector represents correlations with the high-fidelity model (i.e., $\rhovec = \left[\corrop{Q}{Q_1}, \dots, \corrop{Q}{Q_M} \right]^\mathrm{T}$), and $\circ$ denotes the Hadamard (element-by-element) product. The $\fvec$ vector and $\Fmat$ matrix are defined using the cardinality of the sets in $\zvec$ as well as the cardinality of the intersections of those sets. Definitions of $\fvec$ and $\Fmat$ as well as the complete derivation of Equation~\eqref{eq: acv_variance} can be found in the Appendix. The variance reduction of ACV is bounded by the variance reduction of OCV, i.e., $0 \leq R^2_{ACV} < R^2_{OCV} \leq 1$, and it was demonstrated in~\cite{gorodetsky2020generalized} that the main advantage of ACV estimators like ACV-MF and ACV-IS (which are non recursive) is that for $r_i \rightarrow \infty$, $R^2_{ACV} \rightarrow R^2_{OCV}$, while recursive approaches like MLMC and MFMC are limited by the largest correlation within the low-fidelity model ensemble.  Further details are provided in ~\cite{gorodetsky2020generalized}. 

\subsection{Generalized ACV Estimators}
While ACV estimators like ACV-MF and ACV-IS are optimal in the limiting case of $r_i \rightarrow \infty$, recursive estimators can still be advantageous for small $r_i$ regimes. Therefore, the optimal ACV estimator depends on the available data and generalized ACV estimators were introduced in~\cite{bomarito_improving_2022}. These generalized ACV estimators allow for the restructuring of the control variates according to a graph, where the graphs are defined as a series of directed connections 
$i \rightarrow j$ that indicate that the model $Q_i$ acts as a control variate of $Q_j$. Recursive estimators like MLMC and MFMC correspond to an ordered sequence, while ACV estimators like ACV-MF and ACV-IS correspond to a peer structure; generalized ACV estimators 
can encode graphs that simultaneously include ordered sequences and peer structures. From a numerical standpoint, when the number of low-fidelity models is small, the best graph can be selected through enumeration, i.e., all solutions are computed and the one with the smallest variance is selected. The graph effect is represented by a different representation of the vector $\fvec$ and the matrix $\Fmat$; see~\cite{bomarito_improving_2022} for additional details. 
In the next section the resource allocation is presented for any ACV estimator, given a fixed set of models, a prescribed sampling scheme, and a specific graph.

\subsection{The Resource Allocation Problem}
The most important aspect of multifidelity sampling estimators is that computational resources can be optimally allocated across the ensemble of models to minimize the estimator variance for a prescribed computational budget\footnote{An alternate formulation to minimize computational cost given an accuracy constraint on $\Var{\acvest (\zvec)}$ is also common.}. Let the total cost $\mathcal{C}$ of an ACV estimator
\begin{equation}\label{eq:acv_cost}
\mathcal{C}(\zvec) = Nw + \sum_{i=1}^M N_iw_i,
\end{equation}
where $w$ and $w_i$ are the cost of the high-fidelity model and low-fidelity model $i$, respectively. The resource allocation problem is formulated as 
\begin{equation}\label{eq:sample_allocation_optimization}
\zvec^* = \argmin_{\zvec}  \Var{\acvest\left(\hat{\zvec}\right)} \quad \text{s.t.} \quad \mathcal{C}(\zvec)  < C_{budget}, \\
\end{equation}
where $\Var{\acvest\left(\hat{\zvec}\right)}$ is given by Eq.~\eqref{eq: acv_variance} and the optimal solution $\zvec^*$ corresponds to the number of simulations for each model, i.e., $N$ and $N_i = card(\sset{i}{})$ for $i=1,\dots,M$, given an assigned partitioning scheme for $\sset{i}{1} \cup \sset{i}{2} = \sset{i}{}$.

A general solution of this standard ACV sample allocation problem is currently unavailable due to the difficult nature of the optimization.  Current approaches find an optimum on restricted subdomains of $\zvec$ (such as those in Table \ref{tab: summary_estimators_ACV}).  The most general of these approaches, parametric ACVs \cite{BOMARITO2022110882}, considers many such restricted subdomains through parametric construction.  The construction technique is based on developing trees (hierarchies) describing the relations of all sets in $\zvec$ and then determining the recursive relation that minimizes estimator variance. Though subdomains generated in this way do not completely cover the domain of all possible $\zvec$, the increased coverage compared to the methods in Table \ref{tab: summary_estimators_ACV}  typically results in lower variance. 

\subsection{The Model Tuning Problem}
In this section, the main contribution of the paper is introduced. As described above, the key problem in multifidelity modeling is to optimally allocate computational resources across the available models. In this work, this framework is expanded by considering that each low-fidelity approximation (model) $Q_i$ can be a function of hyperparameters $\beta_i$ where $\hyper$ denotes the collection of all model hyperparameters: $\hyper = {\beta_1 \cup \beta_2 \cup \dots \cup \beta_M}$. The high-fidelity model is assumed to be fixed and not tunable in order to provide for an invariant reference model\footnote{Accounting for additional accuracy/cost resulting from high-fidelity tuning would typically require an additional reference, e.g. experimental data, which introduces an additional information source and expands the scope of the ensemble.}.
Furthermore, for simplicity in the following, it is assumed that all low-fidelity models have hyperparameters, but this is not necessary and some low-fidelity models could be fixed. The expectation of $Q$ is approximated with an estimator $\acvest$, i.e., $\EE{Q} \approx \acvest (\zvec,\hyper)$, where now an explicit dependency on the hyperparameter vector is introduced. To capture the effect of hyperparameters on the costs of the low-fidelity models (e.g., $\hyper$ can include a control on the spatial resolution of a low-fidelity model), Eq.~\ref{eq:acv_cost} is modified to include the dependency of hyperparameters,
\begin{equation}\label{eq:acv_cost_beta}
\mathcal{C}(\hyper, \zvec) = Nw + \sum_{i=1}^M N_iw_i(\beta_i).
\end{equation}
The model tuning problem can now be formulated by augmenting the resource allocation problem (Eq.~\eqref{eq:sample_allocation_optimization}) with a simultaneous search for the optimal hyperparameters\footnote{See Sec.~\ref{section:optimization_approach} for additional formulation details.} as follows:
\begin{equation}\label{eq: general_model_tuning_beta}
\zvec^*,\beta^* = \argmin_{\zvec,\beta} \Var{\acvest (\zvec,\hyper)} \quad \text{s.t.} \quad \mathcal{C}(\zvec,\hyper) < C_{budget},
\end{equation}
where the ACV estimator is now also a function of the hyperparameters through the control variates, 
\begin{equation}\label{eq: approximate_CV_beta}
\begin{split}
 \acvest\left(\hyper,\avec,\zvec\right) &= \mcest(\sset{}{}) + \sum_{i=1}^M \alpha_i \left( \mcest_i(\beta_i, \sset{i}{1}) - \mcest_i(\beta_i, \sset{i}{2}) \right) = \mcest(\zvec) + \avec^\mathrm{T} \diffvec(\hyper,\zvec)\\
\end{split}
\end{equation}
and its variance follows directly from Eq.~\ref{eq: acv_variance}, i.e., $\Var{\acvest\left(\zvec ,\hyper \right)} = \Var{\mcest(\sset{}{})} \left(1 - R^2_{ACV}(\hyper) \right)$ with 
\begin{equation}\label{eq: acv_variance_red_beta}
R^2_{ACV}(\hyper) = \left[\fvec(\zvec) \circ \rhovec(\hyper) \right]^\mathrm{T}\left[\Fmat(\zvec) \circ \Pmat(\hyper) \right]^{-1}\left[\fvec(\zvec) \circ \rhovec(\hyper) \right].
\end{equation}

It should be noted that the optimal weights in Eq.~\ref{eq: optimal_alpha} and the vector $\rhovec$ and the matrix $\Pmat$ in Eq.~\ref{eq: acv_variance_red_beta} depend on $\hyper$ since $\rho_i(\beta_i) = \corrop{Q}{Q_i(\beta_i)}$ and $P_{ij}(\beta_i,\beta_j) = \corrop{Q_i(\beta_i)}{Q_j(\beta_j)}$, respectively.

\subsection{Pilot Sampling}

It can be seen from Equation~\eqref{eq: acv_variance_red_beta} that solving Eq.~\ref{eq: general_model_tuning_beta} for the combined ACV sample allocation and model tuning requires knowledge of the dependence of model correlations with respect to the hyperparameters, $\rhovec(\hyper)$ and $\Pmat(\hyper)$. In general, these correlations must be estimated empirically (for fixed hyperparameter values) by drawing $N_{pilot}$ random samples of $\zvec$ and evaluating each model, i.e., by using \textit{pilot sampling}. Additional strategies to capture the dependence on $\beta$ should be introduced, as detailed in the next section.

Historically, there are three different solution modes available in Dakota~\cite{adams2022deployment} for solving ACV sample allocation optimization with respect to pilot sampling: 1) \emph{offline pilot}, 2) \emph{pilot projection}, and 3) \emph{online pilot}. The offline pilot approach treats pilot sampling as an offline cost that precedes an online phase where the available budget is allocated. That is, the initial cost of evaluating the pilot samples is excluded from the budget. Note that this is the most commonly used approach in the multifidelity uncertainty propagation literature \cite[e.g.,][]{gorodetsky2020generalized,peherstorfer2016optimal,BOMARITO2022110882}. This mode can be employed to obtain an ``oracle'' reference solution since it provides access to a potentially fully converged set of correlations/covariances, without impacting the available computational resources within $C_{budget}$. 

Pilot projection, by contrast, is an online mode that assumes that the pilot samples will be reused and thus includes their cost within the overall computational budget. To do so, the sample allocation optimization problem is constrained to have the pilot investment as a lower bound. Typically, $N_{pilot}$ is chosen such that the pilot evaluation is a modest proportion of the overall computational budget, and then the performance of an ACV estimator is projected based on the initial model correlation estimates.  That is, the estimator variance is minimized over $\{N,N_i\}$, but the pilot-based correlations are not updated, the additional sample evaluations are not carried out, and therefore the final moment statistics are not computed. 

Online pilot mode, or iterated ACV 
\cite{dakota2024,adams2022deployment},  
 begins in the same manner as pilot projection, using a (typically small) initial pilot evaluation to project estimator performance. However, this approach then aims to iterate towards the optimal sample allocation in term of the number of shared samples evaluated for all models, since these sample increments can be used to update the correlation estimates. 
 The iteration proceeds by repeating the following steps: 1) compute the optimal sample allocation given the current estimate of model correlations, 2) calculate the change in number of shared samples, $\Delta N$, between the current and previous sample allocation, 3) evaluate a new $\Delta N$ pilot samples\footnote{In practice, initial $\Delta N$ estimates may be damped using an under-relaxation parameter to mitigate potential over-estimation of shared samples when model correlations are inaccurate.} and update the estimated model correlations. The process stops when $\Delta N \leq 0$, then transitioning to the evaluation of low-fidelity sample increments $N_i$ followed by the roll-up of final statistics.

Each of these three solution modes are useful in different settings. Offline pilot mode is commonly used to generate reference or oracle solutions that result from (approximately) exact \textit{a priori} knowledge of model statistics. Pilot projection mode is helpful for evaluating the performance of different ACV estimators or for selecting the best ensemble of models. Online pilot mode has the advantage of balancing accuracy and cost in an online analysis, helping to avoid under-estimating $N_{pilot}$ (incurring reduced accuracy) or over-estimating $N_{pilot}$ (incurring reduced efficiency). Note that multiple approaches can be combined in a single analysis, e.g., using pilot projection to select the best ACV method and then online pilot mode to iterate with this selection. As shown later, all three solution modes are used for different purposes in this work\footnote{
Starting with version 6.20, Dakota adds a fourth solution mode for offline projection, augmenting the existing (online) pilot projection in order to support all combinations of online vs. offline and pilot vs. projection. This mode will not be included in the studies that follow.}.

\subsection{Optimization Approach}\label{section:optimization_approach}

The numerical solution of the optimization problem described in Equation~\eqref{eq: general_model_tuning_beta} is now discussed: this equation illustrates that model tuning (optimizing $\hyper$) is tied with sample allocation optimization (optimizing $\zvec$). Two solutions strategies can be designed for this task~\cite{adams2022deployment}, namely an \emph{all-at-once} or a \emph{bi-level} strategy. These two strategies offer different trade-offs, and are described below. In the case of an all-at-once optimization, a surrogate for each term that depends on $\hyper$ is needed. In particular, since the dependency of $\Var{\acvest\left(\zvec ,\hyper \right)}$ on $\hyper$ (Equation~\eqref{eq: general_model_tuning_beta}) is encapsulated in Equation\ref{eq: acv_variance_red_beta}, it is necessary to obtain a surrogate for each term of $\rhovec$ and $\Pmat$ which, accounting for the symmetry in $\Pmat$, corresponds to a total number of $M(M+1)/2$ surrogates. Moreover, for each of the $M$ low-fidelity models, a surrogate to represent the dependency of its cost $w_i$ on $\beta_i$ is needed. Each of these surrogates can potentially have a different dimension since each $\beta_i$ vector has a dimension associated with the hyperparameters of the $i$th model. To build these surrogates, the pilot sampling solution described in the previous section can be extended to a set of points in the support of $\hyper$, e.g., by using a regular tessellation. At each location $\hyper^{b}$ of the tesselation of the hyperparameter space, a standard pilot sampling procedure is carried out and all the required statistics are evaluated, i.e., $\rhovec(\hyper^{b})$, $\Pmat(\hyper^{b})$, and $w_i(\hyper^{b})$. Surrogates are then built from $N_b$ realizations of these quantities $\rhovec(\hyper^{b})$, $\Pmat(\hyper^{b})$, and $w_i(\hyper^{b})$ for $b=1, \dots, N_b$ via standard techniques like regression or interpolation. Once the surrogates are trained, Equation~\eqref{eq: general_model_tuning_beta} can be solved by resorting to standard solutions for the multifidelity resources allocations. For instance, an example of this strategy based on local quadratic polynomials and Sequential Least Squares Programming (SLSQP) was presented in~\cite{bomarito_improving_2022}. It is easy to see that, depending on the number of hyperparameters per model and the number of models, this approach may be difficult to efficiently scale.

In this work, a bi-level optimization strategy is chosen to solve Equation \eqref{eq: general_model_tuning_beta} where $\hyper$ is optimized on the outer loop with the inner loop performing sample allocation optimization for each instance of $\hyper$:
\begin{equation}\label{eq:bi_level_tuning}
\hyper^* = \argmin_{\hyper} \left[  \min_{\zvec}  \Var{\acvest\left(\hyper,\zvec \right)} \quad \text{s.t.} \quad \mathcal{C}(\hyper, \zvec)  < C_{budget} \right]. 
\end{equation}
Here, it is emphasized that the inner loop optimization problem depends on $\hyper$ through the required estimates of model correlations $\rhovec(\hyper)$ and $\Pmat(\hyper)$. The bi-level approach decouples the model tuning problem, allowing different off-the-shelf optimization algorithms for the outer level to be combined with varying ACV estimator schemes (Table \ref{tab: summary_estimators_ACV}) on the inner level. This includes the parametric ACV approach for an additional search over model graphs on the inner level as demonstrated later on in the manuscript. Furthermore, the outer level optimization strategy only requires a single surrogate for the $\Var{\acvest\left(\hyper,\zvec \right)}$ resulting from the inner level optimization, rather than a surrogate for each term of $\rhovec(\hyper)$ and $\Pmat(\hyper)$ as in the case of the all-at-once optimization. 

The primary challenge of the bi-level approach is that the outer-loop objective function is noisy since $ \Var{\acvest}$ is evaluated from estimates of $\rhovec(\hyper)$ and $\Pmat(\hyper)$  based on finite (and often limited) pilot sampling. Note that either online pilot or pilot projection can be used to solve the inner-level sample allocation optimization depending on whether the accuracy or efficiency of the solution will be prioritized, respectively. Pilot projection is adopted in this work in order to avoid over-investment in the model tuning process, since only the samples for $\hyper^*$ can be reused.

For the outer loop, the efficient global optimization (EGO) algorithm \cite{jones1998efficient} is used for its ability to handle the effects of numerical noise while minimizing $ \Var{\acvest}$. EGO is a global surrogate-based optimizer that approximates the objective function with a Gaussian process (GP) response surface (i.e., a \textit{Bayesian optimization} technique). EGO fits an initial GP to objective function values evaluated at $N_{init}$ random candidate points\footnote{This work uses $N_{init} = (N_{\beta}+1)(N_{\beta}+2)/2$ for all examples, where $N_{\beta}=| \hyper |$ is the total number of tunable hyperparameters.} and then iteratively refines the GP using points selected from the maximization of an expected improvement (EI) acquisition function. A strength of EGO is its ability to select the next candidate point in such a way that balances the need to minimize the approximate objective (exploitation) with the need to improve the approximation where it is most uncertain (exploration). Algorithm \ref{alg:EGO} summarizes the use of EGO for solving the model tuning optimization problem.Note that while EGO is an effective choice for low-dimensional model tuning problems like those presented in Section \ref{section:results}, a more scalable optimization algorithm (e.g., trust-region model management \cite{trmm_ref}) is recommended for locally-optimal solution to higher dimensional problems \cite{adams2022deployment}.

In this work, the portion of the overall computational budget, $C_{budget}$, dedicated to model tuning is loosely controlled by specifying the number of pilot samples, $N_{pilot}$, and maximum number of iterations, $N_{iter}$ for the EGO algorithm. It is important to point out that a potentially significant portion of computation performed during model tuning can be reused in a subsequent multifidelity uncertainty propagation process. For example, $N_{pilot}$ evaluations of the high-fidelity model can be saved and reused as well as the evaluations of the low-fidelity models at the optimal hyperparameters, $\hyper^*$. Note that using a convergence tolerance that terminates EGO based on diminishing returns from model tuning could be used in lieu of (or in addition to) setting  $N_{iter}$. Doing so would have some practical advantages, but was not pursued here and instead left for future studies.

The actual cost of model tuning that cannot be reused and is thus viewed as an overhead will be denoted by $C_{\mathcal{T}}$. To this end, let $\mathcal{T}_{\hyper}$ denote the set of hyperparameter values that were considered by the outer loop optimizer \textit{excluding} the optimal value returned:
\begin{equation}\label{eq:non_optimal_hyper}
	\mathcal{T}_{\hyper} = \{ \hyper^{(j)} \}_{j=1}^{N_{iter}} \setminus \hyper^*.
\end{equation}
Then, the model tuning overhead can be calculated as follows
\begin{equation}
	 C_{\mathcal{T}} = \sum_{ \hyper \in \mathcal{T}_{\hyper}} \sum_{i=1}^M N_{pilot} w_i(\beta_i) \label{tuning_overhead}
\end{equation}
where it is again assumed that all $M$ low-fidelity models are tunable. Thus, when subsequently constructing the ACV estimator, the overall computational budget is modified as $C_{budget} - C_{\mathcal{T}}$.

\begin{algorithm}
\caption{Efficient Global Optimization (EGO) for Model Tuning}
\begin{algorithmic}[1]
\State \textbf{Input:} Number of initial candidate points $N_{init}$, number of iterations $N_{iter}$, number of pilot samples $N_{pilot}$, computational budget $C_{budget}$
\State \textbf{Let:} $J(\hyper) \equiv \min_{\zvec} \Var{\acvest(\hyper, \zvec)} \quad \text{s.t.} \quad \mathcal{C}(\hyper, \zvec) < C_{budget}$ \Comment{EGO objective function}

    \State Randomly select $N_{init}$ hyperparameter candidate values $\{\hyper_i\}_{i=1}^{N_{init}}$
    \State Initialize dataset: $\mathcal{D}_0 = \{\hyper_i, J(\hyper_i)\}_{i=1}^{N_{init}}$ \Comment{Evaluate $J(\hyper)$ for each candidate}

\For{$t = 1$ to $N_{iter}$}
    \State Fit Gaussian Process surrogate model to $\mathcal{D}_{t-1}$
    \State Select candidate point $\hyper_t$ to maximize the expected improvement in $J$
    \State Estimate $\rhovec(\hyper_t)$ and $\Pmat(\hyper_t)$ with $N_{pilot}$ pilot samples
    \State Update  dataset: $\mathcal{D}_t = \mathcal{D}_{t-1} \cup \{\hyper_t, J(\hyper_t)\}$
\EndFor

\State \textbf{Return:} Optimal hyperparameters $\hyper^* = \arg\min_{(\hyper_i, J(\hyper_i)) \in \mathcal{D}_{N_{iter}}} J(\hyper_i)$
\end{algorithmic}  \label{alg:EGO}
\end{algorithm}

\subsection{Summary: Online  Multifidelity Estimators with  Model Tuning and Iterated ACV}

\begin{figure}
  \centering
  \includegraphics[width=\textwidth]{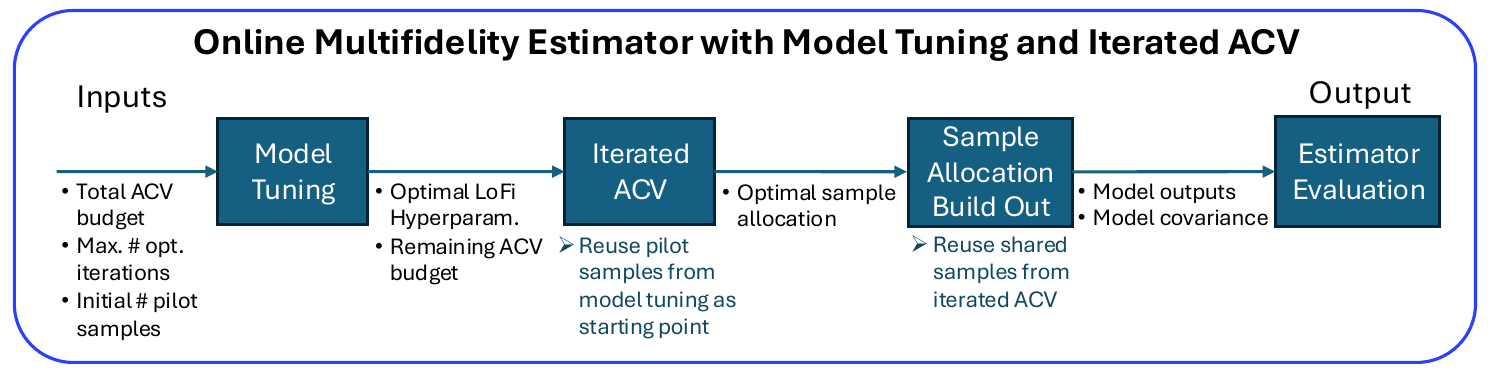}
  \caption{The online procedure for constructing a multifidelity estimator using automated model tuning and iterated ACV used in this work.}
  \label{fig:model_tuning_overview}
\end{figure}

Figure \ref{fig:model_tuning_overview} summarizes the overall procedure for obtaining ACV estimators, $\acvest\left(\hyper,\avec,\zvec\right)$, via Equation \eqref{eq: approximate_CV_beta} in the practical setting where neither optimal low-fidelity model settings ($\hyper$) nor knowledge of model statistics ($\Pmat$, $\rhovec$) are available \textit{a priori}. The diagram highlights two instances where computations can be reused from the previous step to improve efficiency: 1) Iterated ACV reuses pilot samples from the model tuning step to form initial estimates of $\Pmat$ and $\rhovec$ that are then iterated upon, and 2) the sample allocation build out starts from the final collection of shared pilot samples from the iterated ACV step, only needing to evaluate the low-fidelity models at samples not shared across all models. See Algorithm \ref{alg:summary} for  more details.

Note that larger values of $N_{pilot}$ and $N_{iter}$ will tend to lead to more accurate model tuning optimization solutions, but will also incur a larger computational overhead, $C_{\mathcal{T}}$, leaving less budget devoted to reducing the variance of the resulting ACV estimator. It is this cost-versus-precision tradeoff that depends on $N_{pilot}$, $N_{iter}$, $C_{budget}$, and the application-dependent complexity of the model tuning optimization problem that is the focus on this study.

\begin{algorithm}
\caption{Online Multifidelity Estimator with Model Tuning and Iterated ACV}
\begin{algorithmic}[1]
\State \textbf{Input:} Computational budget $C_{budget}$, number of initial candidate hyperparameter values $N_{init}$, number of optimization iterations $N_{iter}$, number of (initial) pilot samples $N_{pilot}$, bounds for low-fidelity hyperparameters $\hyper$ 

\Statex
\State \textbf{Step 1: Solve model tuning optimization problem}  \Comment{Via Algorithm \ref{alg:EGO}}
    \State \;\;\;\; \textbf{Input:} $N_{iter}$, $N_{pilot}$, $N_{init}$, bounds for $\hyper$ 
    \State \;\;\;\; \textbf{Return:} Optimal hyperparameters $\hyper^*$, model tuning overhead $C_{\mathcal{T}}$

\Statex
\State \textbf{Step 2: Perform iterated ACV sample allocation optimization}
    \State \;\;\;\; \textbf{Input:} $\hyper^*$, remaining budget $C_{budget} - C_{\mathcal{T}}$
    \State \;\;\;\; \textbf{Return:} Optimal sample allocation $\zvec^*$, estimated model statistics $\{ \Pmat, \rhovec \}$,  estimated variance $\Var{\acvest(\hyper^*, \zvec^*)}$

\Statex
\State \textbf{Step 3: Evaluate models according to optimal sample allocation}
    \State \;\;\;\; \textbf{Input:} $\hyper^*$,  $\zvec^*$
    \State \;\;\;\; \textbf{Return:} Outputs for all models evaluated on $\zvec^*$

\Statex
\State \textbf{Step 4: Form ACV estimator}  \Comment{Via Equation \eqref{eq: approximate_CV_beta} }
    \State \;\;\;\; \textbf{Input:} Sample allocation model outputs, model statistics $\{ \Pmat, \rhovec \}$
    \State \;\;\;\; \textbf{Return:}  ACV estimator $\acvest$

\Statex
\State \textbf{Return:} Optimal hyperparameters $\hyper^*$, ACV estimator $\acvest$, estimated variance $\Var{\acvest(\hyper^*, \zvec^*)}$
\end{algorithmic} \label{alg:summary} 
\end{algorithm}

It is worth highlighting that most existing work assesses ACV estimator performance in terms of estimated variance via Equation \eqref{eq: acv_variance}, requiring only steps 1 and 2 in Algorithm \ref{alg:summary} to be executed . Furthermore, \textit{a priori} knowledge of $\Pmat$ and  $\rhovec$  is typically assumed using \textit{offline pilot} solution mode, ignoring the errors and costs associated with pilot sampling. By contrast, carrying out the full procedure presented here allows performance to be quantified in terms of mean squared error (MSE)  on the final ACV estimator while taking into account model tuning and pilot sampling costs. Assuming a reference solution, $\mcest^{ref}$,  is available (e.g., via brute force Monte Carlo estimation), the MSE can be calculated as follows
\begin{equation}\label{eq:mse}
	MSE = \frac{1}{N_{trial}} \sum_{j=1}^{N_{trial}} \left(\acvest^{(j)} - \mcest^{ref} \right)^2,
\end{equation}
where $\acvest^{(j)}, j = 1,...,N_{trial} $ are ACV estimators generated by carrying out the procedure above for $N_{trial}$ random trials.

%
%
%


\section{Application}

This work demonstrates the practical aspects of automated model tuning for multifidelity uncertainty propagation in the context of trajectory simulation for entry, descent, and landing (EDL). Onboard trajectory simulation is emblematic of problems where little \textit{a priori} knowledge about optimal low-fidelity model settings and model statistics (e.g., correlations) may be available and computational efficiency is critical. In order to be practically useful then, multifidelity estimators must be constructed in an online setting where upfront model tuning and pilot sampling costs cannot be ignored as they often are in existing literature.

\begin{figure}
    \centering
    \includegraphics[width=0.8\textwidth]{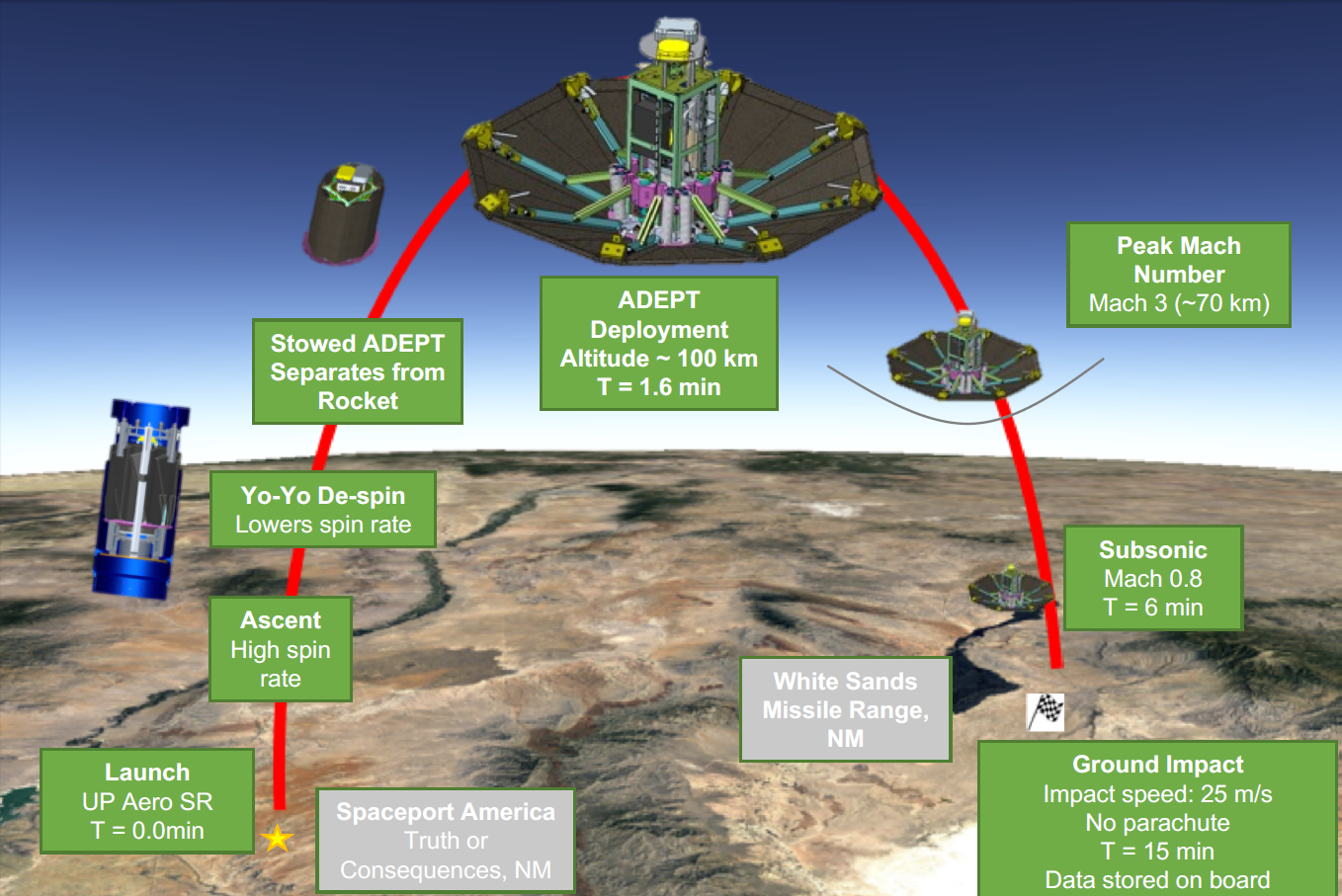}
    \caption{Schematic of the ADEPT SR-1 concept of operations \cite{dutta2019flight}}
    \label{fig: adept}
\end{figure}

The particular problem considered here is the trajectory simulation of a flight test conducted on an atmospheric entry system called the Adaptable Deployable Entry and Placement Technology (ADEPT) \cite{cassell2018adept}. ADEPT is an umbrella-like heatshield made of 3-D woven carbon fabric that acts as both a thermal protection system and a structural membrane.  In smaller sizes, ADEPT can be equipped on small satellites for direct entry and/or aerocapture.  At larger sizes it could enable human-scale missions with a large drag area with reduced launch sizes and weights.  The ADEPT Sounding Rocket One (ADEPT SR-1) flight test was performed to test its ability to deploy exo-atmospherically into the specified aerodynamic configuration  as well as to evaluate the aerodynamics of its first-of-a-kind faceted geometry  (See Figure \ref{fig: adept}). Likewise, the novelty of this test flight has led to ADEPT SR-1 becoming a testbed for trajectory simulation and uncertainty propagation techniques.

Trajectory simulation was originally performed for the ADEPT flight test using a traditional MC approach and compared favorably to flight reconstructions from onboard measurement systems \cite{dutta2019flight}.  NASA Langley Research Center's POST2 software is used for the 6-degree-of-freedom trajectory simulation of ADEPT SR-1; the simulations are based on aerodynamics from both computational fluid dynamics simulations and wind tunnel test data.  At its core, POST2 performs a numerical integration of the equations of motion; it is event-driven, but also relies on integration given a time step parameter $\Delta t$. A time step of $\Delta t=0.001$ was used for the ADEPT-SR simulation \cite{dutta2019flight}. 

Ultimately, the trajectory simulation gives a means for predicting 15 QoIs describing the flight state given a sample of 76 uncertain input parameters (i.e., $Z \in \mathbb{R}^{76}$ in Equation \eqref{eq: approximate_CV}) including those describing the deployment orientation and atmospheric conditions.  Though the MC method was shown to produce reasonable results, the analysis was revisited with a multifidelity uncertainty propagation approach to significantly improve computational efficiency \cite{warner_multi-model_2021}.  Four models were used in that work: (1) the high-fidelity POST2 model used in the original MC-based work, (2) a POST2 model with reduced physics, i.e., simplified atmospheric model, (3) a POST2 model with coarse timestep, i.e., reduced integration accuracy, and (4) a machine learning surrogate model where a support vector machine was trained on samples of the high-fidelity model.  The models had varying levels of computational speedup compared to the high-fidelity model, as seen in Table \ref{tab: models_table}. Here, it is assumed  that the training of the machine learning surrogate model (including generation of training data) is an offline cost and is thus not factored into the speedup value. Notably, the reduced physics and coarse timestep also had hyperparameters ($\Delta t$) which were hand-selected with best engineering judgment. This work focuses on exploring the benefits of automated model tuning of the time step hyperparameters relative to the performance of using hand tuned values.

\begin{table}[h!]
  \small
  \centering
  \caption{Models considered for multifidelity uncertainty propagation.}
  \label{tab: models_table}
  \begin{tabular}{p{2.7cm}p{2.3cm}p{1.2cm}p{2.75cm}p{1.3cm}}
    \hline \hline
    \textbf{Model}  & \multicolumn{2}{c}{\textbf{Hand Selected \cite{warner_multi-model_2021}}} & \multicolumn{2}{c}{\textbf{Tuned}} \\ 
    & Hyperparameter & Speedup  & Hyperparameter & Speedup \\
    \hline
    High Fidelity & $\Delta t = 0.001 $ & 1X  &  Fixed: $\Delta t = 0.001 $   & 1X  \\
    Reduced Physics & $\Delta t = 0.001$  & 5X  & $\mathbf{\Delta t \in [0.001, 0.25]} $  & [5, 351]X \\
    Coarse Timestep & $\Delta t = 0.1$  & 78X & $\mathbf{\Delta t \in [0.001, 0.25]}$    & [1, 159]X \\
    Machine Learning & -  & 310000X & -    & 310000X \\
    \hline \hline
  \end{tabular}
\end{table}

\begin{figure}
    \centering
    \includegraphics[width=1.5in]{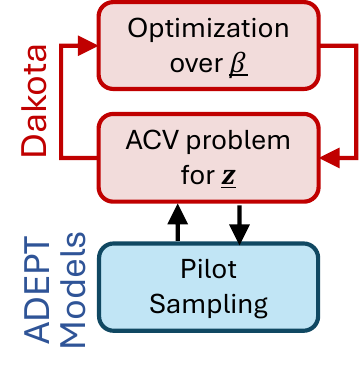}
    \caption{Implementation overview of the model tuning approach. }
    \label{fig:implementation_overview}
\end{figure}

The proposed approach was implemented using Sandia National Laboratories' uncertainty quantification code, Dakota. Figure \ref{fig:implementation_overview} shows a diagram illustrating the implementation approach. Dakota controls the solution of the bi-level model tuning optimization problem (Equation \eqref{eq:bi_level_tuning}), using global surrogate-based optimization (i.e., the EGO algorithm) on the outer loop, passing values of $\hyper$ to ACV sample allocation optimization at each iteration. To solve the ACV problem, calls are dispatched to high- and low-fidelity models for pilot sampling to estimate the required model statistics for calculating estimator variance.

\section{Results}\label{section:results}

This section presents results from two numerical examples that apply model tuning for multifidelity uncertainty propagation in the context of the ADEPT application. The first example focuses on tuning the time step of the \textit{Coarse Time Step}  model (Table \ref{tab: models_table}) only. In this simpler 1D setting, the end-to-end, online uncertainty propagation process depicted in Figure \ref{fig:model_tuning_overview} is carried out to evaluate the performance of the approach in a realistic setting where pilot sampling costs and errors are accounted for in the construction of multifidelity estimators.

The second example considers multiple tuning parameters, optimizing the time step of both the \textit{Reduced Physics}  and \textit{Coarse Time Step} models simultaneously. Here, the focus is on the model tuning optimization solution itself, and performance is gauged based on predicted estimator variance rather than the error in actual multifidelity estimators. By bypassing the evaluation of the sample allocation profiles, it is possible to study performance trends as a function of overall computational budget and to consider multiple trajectory QoIs without incurring an exorbitant computational cost.

Both examples serve to explore the practical benefits of model tuning for a realistic problem and assess the tradeoff of estimator cost versus precision of spending a portion of the overall computational budget on hyperparameter tuning. The performance of multifidelity estimators with model tuning will be benchmarked against three solution types:
\begin{enumerate}
	\item \textbf{Monte Carlo} - exhausting the entire computational budget on sampling the high-fidelity model alone.
	\item \textbf{Hand Selected} - multifidelity estimators generated with fixed models borrowed from previous work \cite{warner_multi-model_2021} that were selected with engineering judgement (Table \ref{tab: models_table}).  For example 1, \textit{online pilot} solution mode will be used to estimate model statistics on the fly during the sample allocation optimization.
	\item \textbf{Best Case} - multifidelity  estimators generated with \textit{a priori} knowledge of both optimal hyperparameter values (no tuning cost) and model statistics (\textit{offline pilot} solution mode with $N_{pilot}=10k$).
\end{enumerate}
The ability of model tuning  to improve upon the Hand Selected solution and approach the Best Case is studied for various settings of $N_{pilot}$, $N_{iter}$, and $C_{budget}$.

\begin{figure} 
\centering 
\subfigure[]{\includegraphics[width=2.9in]{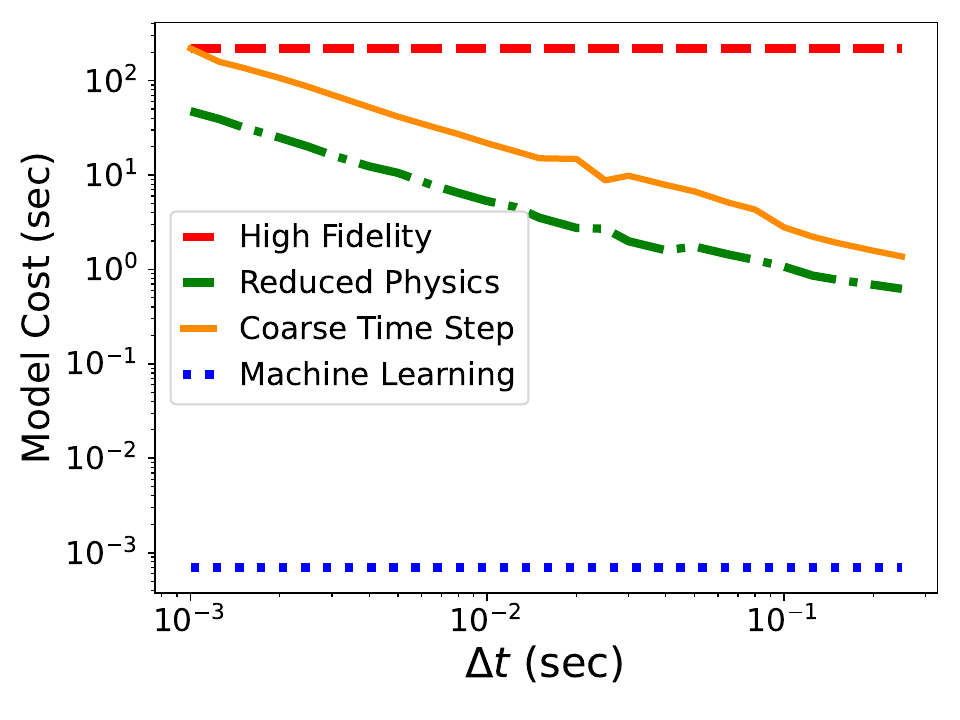}}
\subfigure[]{\includegraphics[width=2.9in]{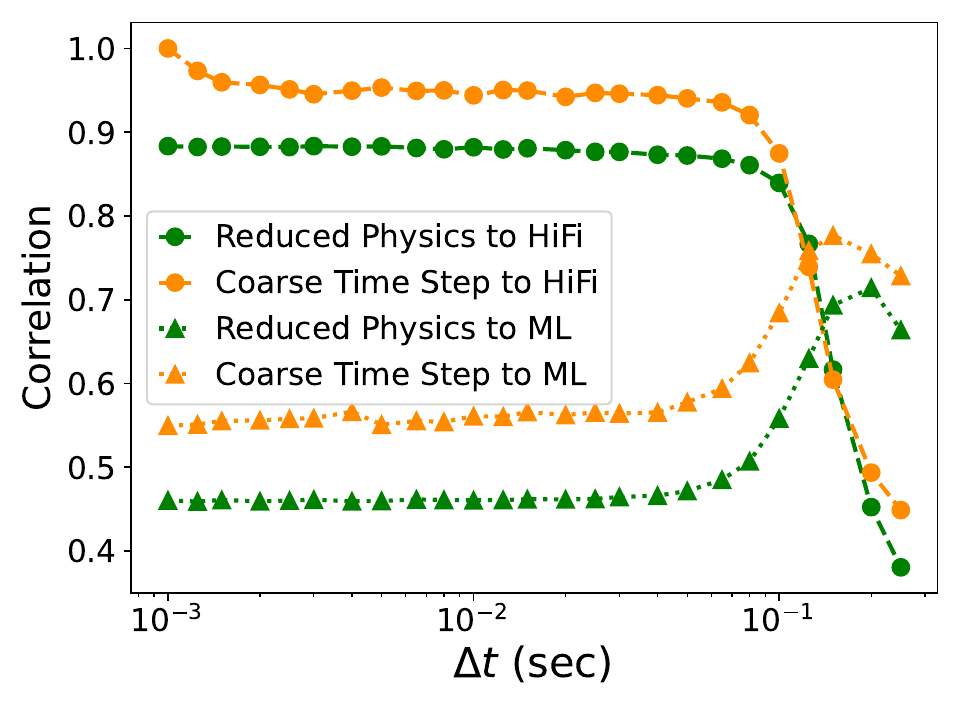}}
\caption{Model costs and correlations versus time step for the time of flight QoI.} \label{fig:model_costs_correlations}
\end{figure}

\begin{figure} 
\centering 
\subfigure[]{\includegraphics[width=2.9in]{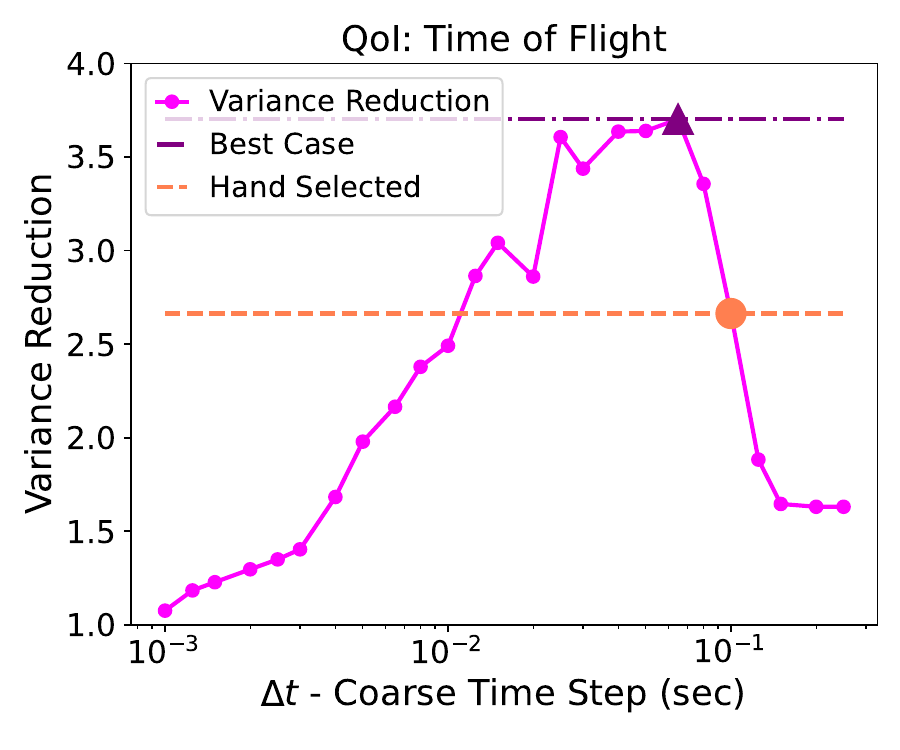}}
\subfigure[]{\includegraphics[width=2.9in]{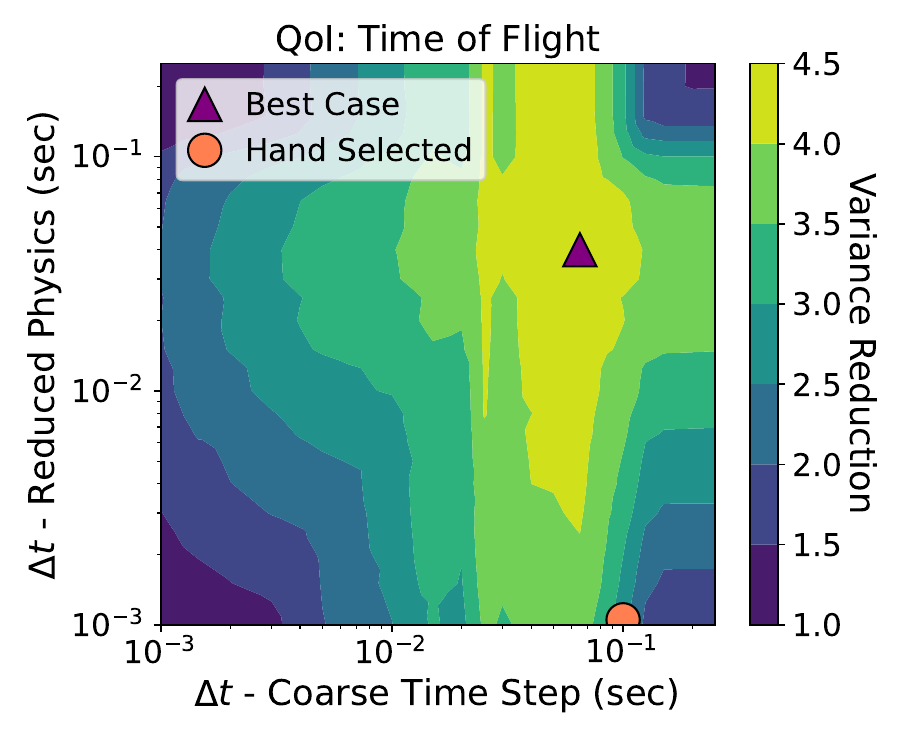}}
\subfigure[]{\includegraphics[width=2.9in]{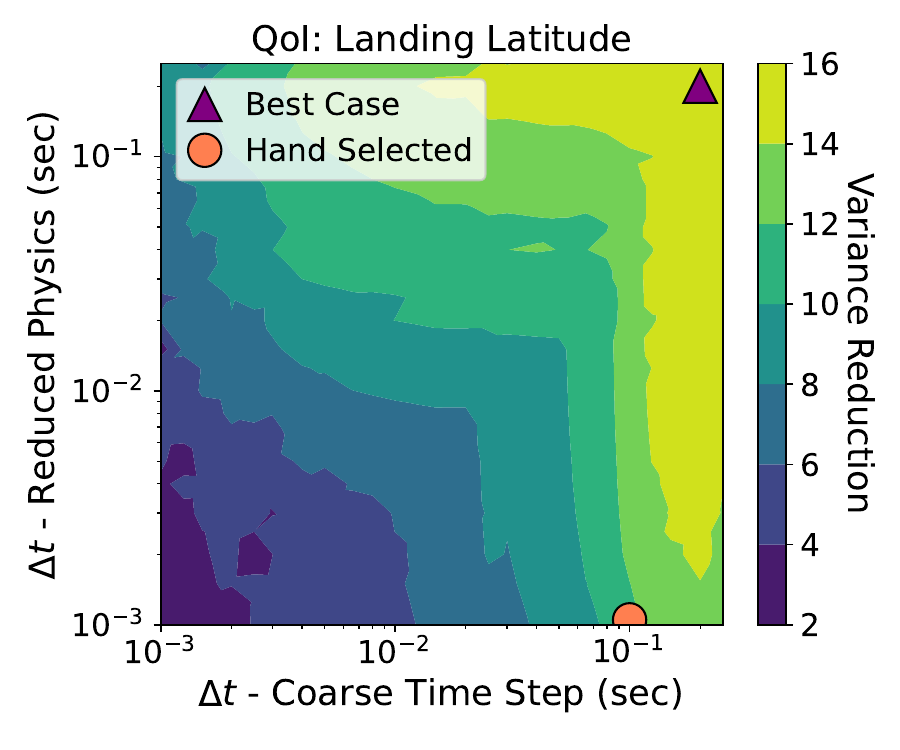}}
\subfigure[]{\includegraphics[width=2.9in]{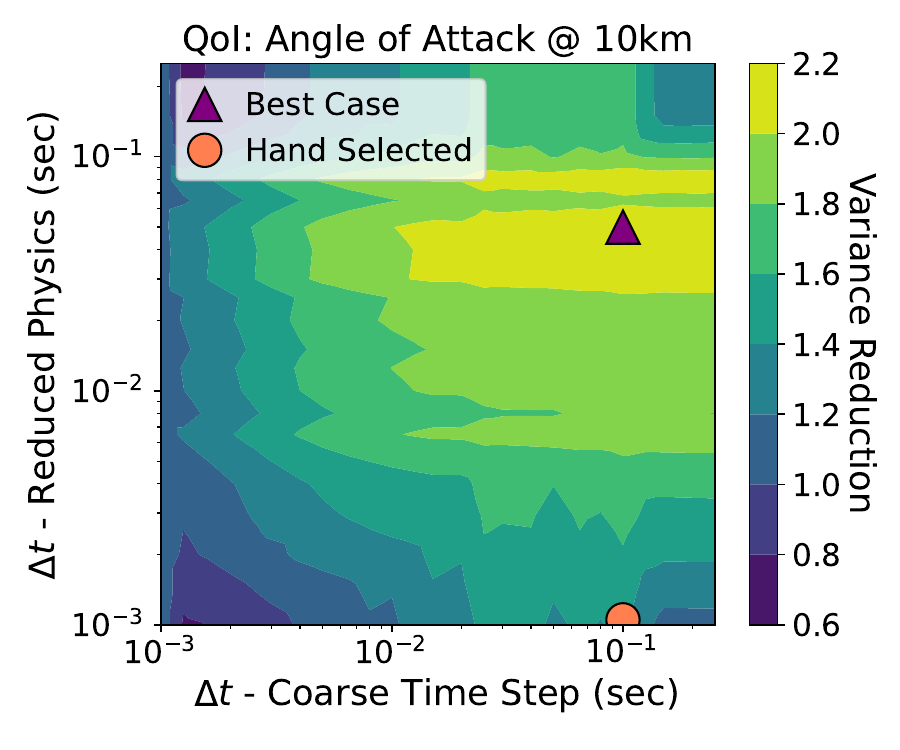}}
\caption{Variance reduction as a function of low-fidelity model time steps for the QoIs considered in example 1 (a) and example 2 (b, c, d).} \label{fig:variance_reduction_contours}
\end{figure}

Figure \ref{fig:model_costs_correlations} illustrates the effect of the tunable time step parameters on the low-fidelity models' computational cost and correlations versus the other models for the time of flight QoI. While the cost of both the \textit{Reduced Physics}  and \textit{Coarse Time Step} models decrease consistently with increasing time step as expected, the model correlations exhibit interesting and non-intuitive behavior. The correlations of both tunable models with respect to the \text{High Fidelity}  and \textit{Machine Learning} models exhibit relatively constant correlation over a large range of increasing time step, but show a drastic decrease/increase with respect to the \textit{High Fidelity}/\textit{Machine Learning} around $\Delta t = 10^{-1}$ sec. The complex nature of this cost versus correlation trade-off makes it difficult to make \textit{a priori} assumptions about valid hand tuned time step values. Note that ACV methods also take advantage of the correlations amongst the low-fidelity models themselves, further complicating manual hyperparameter tuning.

Figure \ref{fig:variance_reduction_contours} shows the variance reduction versus Monte Carlo,
\begin{equation}\label{eq:variance_reduction}
	\text{Variance Reduction} = \frac{ \Var{\mcest}}{ \Var{\acvest}}, 
\end{equation}
 that is possible with the ACV-MF estimator as a function of the tunable time step parameters for a computational budget of 1000 equivalent high-fidelity model evaluations. Here, each time step parameter was discretized as a grid of 25 equally (log) spaced points and the estimator variance, $\Var{\acvest}$, was evaluated at each point using \textit{offline pilot} solution mode with $N_{pilot}=10k$.  Example 1 is a 1D model tuning problem focused on estimating the time of flight QoI (Figure \ref{fig:variance_reduction_contours} (a)) while example 2 is a 2D model tuning problem considering the time of flight, landing latitude, and angle of attack at 10km QoIs (Figure \ref{fig:variance_reduction_contours} (b)-(d)). These plots help to illustrate the nature of the model tuning optimization problem that must be solved by providing a visualization of the objective function to be minimized with EGO. The plots also show the time step values for the \textbf{Hand Selected} and \textbf{Best Case} solutions, highlighting the large difference between those settings as well as the resulting variance reduction. In particular, the optimal time step setting for the \textit{Reduced Physics} model is significantly larger than the hand selected value for all three QoIs in Figures \ref{fig:variance_reduction_contours} b-d. This indicates that there may be a tendency by practitioners to overestimate how accurate a low-fidelity model must be to be useful within a multifidelity framework, where models with large absolute errors can still be valuable provided they are adequately correlated.

\subsection{Example 1 - Single Tunable Hyperparameter}

Example 1 carries out the full online multifidelity estimator procedure with model tuning and iterated ACV shown in Figure \ref{fig:model_tuning_overview}, using the bilevel optimization strategy in Equation \eqref{eq:bi_level_tuning} to select an optimal timestep for the Coarse Time Step model. Estimators for the time of flight QoI are formed using different numbers of pilot samples, $N_{pilot} = \{10, 50, 100\}$, and EGO iterations, $N_{iter}=\{5, 10, 20\}$\footnote{EGO uses an initial sample size $N_{init} = 3$  to generate a surrogate model, so these values represent 2, 7, and 17 EGO iterations beyond the initial sample, respectively}, to solve the model tuning optimization problem. One hundred random trials are performed for each model tuning estimator as well as the \textbf{Monte Carlo}, \textbf{Hand Selected}, and \textbf{Best Case} solution types described above for three different computational budgets, $C_{budget} = \{500, 1000, 2000\}$, defined in terms of equivalent high-fidelity model cost.  Note that the \textbf{Best Case} solution uses an optimal time step value of $\Delta t = 0.065\;\mathrm{sec}$  and the  \textbf{Hand Selected} case uses $\Delta t = 0.1\;\mathrm{sec}$ (see Figure \ref{fig:variance_reduction_contours} a). Each solution approach is compared in terms of the MSE defined in Equation \eqref{eq:mse} with a reference solution computed using Monte Carlo with $N=100k$ samples. For all ACV solutions, the ACV-MF estimator (Table \ref{tab: summary_estimators_ACV})  is used.


\begin{figure} 
\centering 
\subfigure[]{\includegraphics[width=\textwidth]{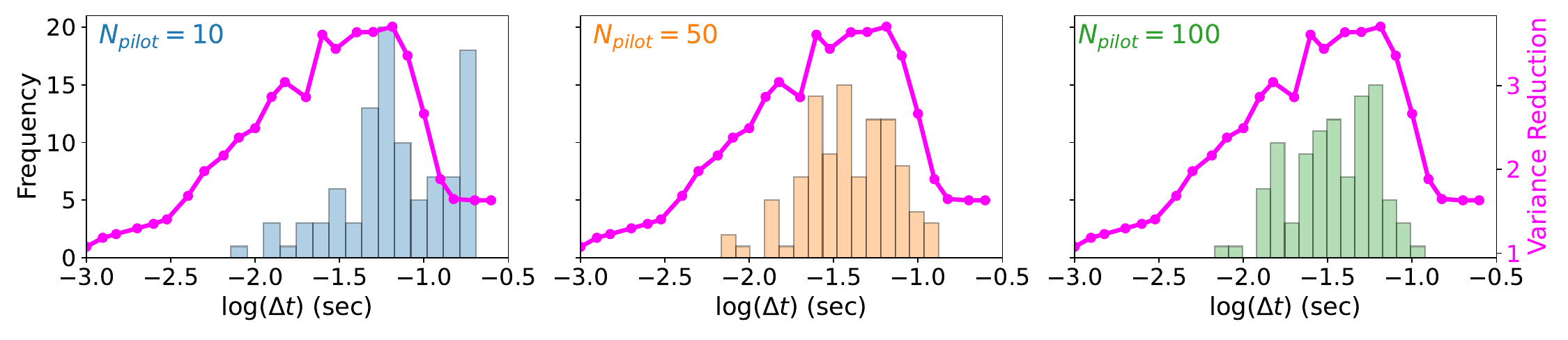}}
\subfigure[]{\includegraphics[width=\textwidth]{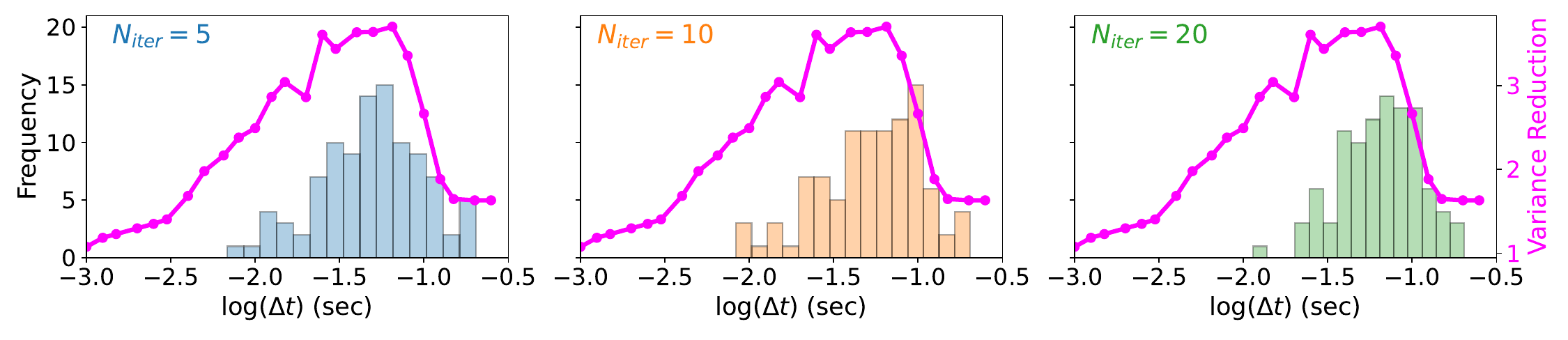}}
\caption{Optimal time steps found across 100 random trials for example 1 for different numbers of a) pilot samples and b) EGO iterations.}
\label{fig:1d_tuning_results}
\end{figure}

\begin{figure} 
\centering 
\subfigure[]{\includegraphics[width=2.9in]{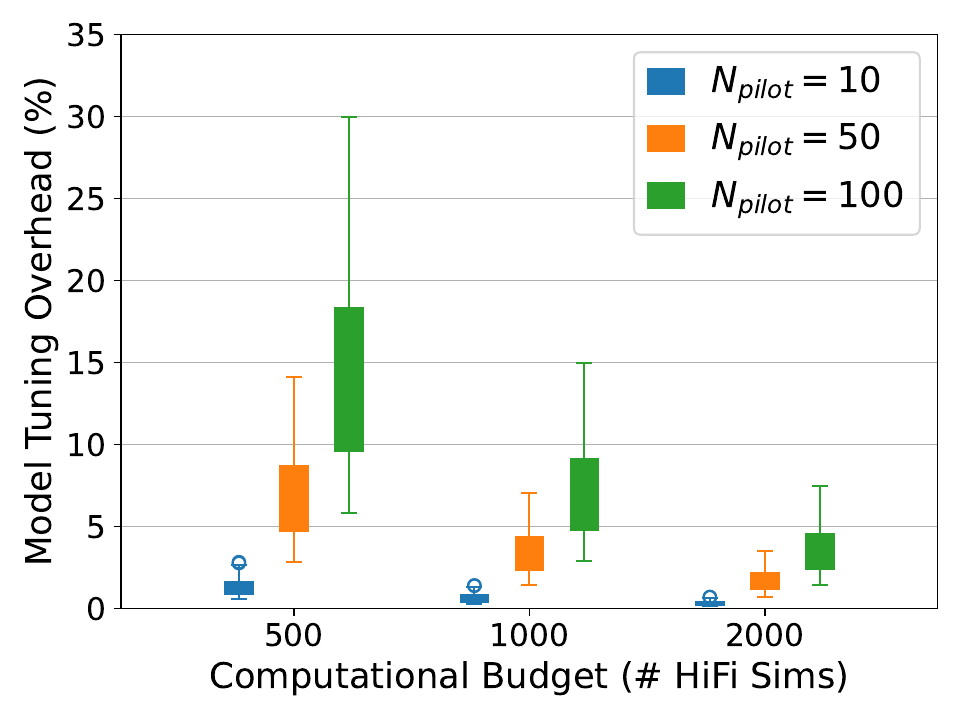}}
\subfigure[]{\includegraphics[width=2.9in]{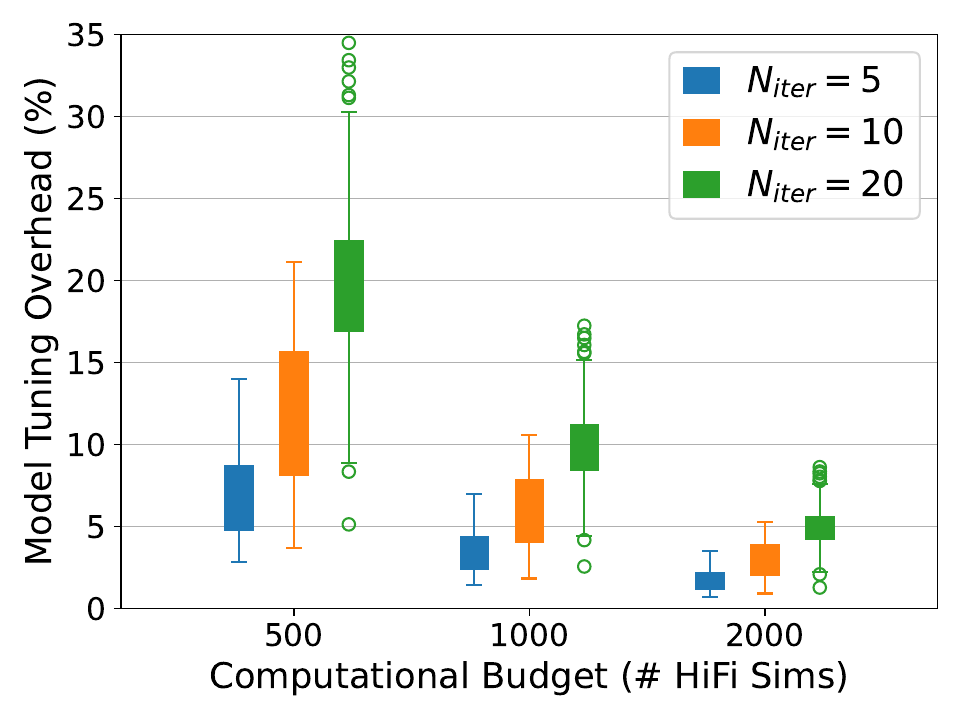}}
\caption{The computational overhead ($\%$) from solving the model tuning optimization problem in example 1 for different numbers of a) pilot samples and b) EGO iterations.}
\label{fig:1d_tuning_overhead}
\end{figure}

Results for the solution of the model tuning optimization problem (Equation \eqref{eq:bi_level_tuning}) are shown in Figures \ref{fig:1d_tuning_results} and \ref{fig:1d_tuning_overhead}. Figure \ref{fig:1d_tuning_results} shows the optimal time step values returned by the EGO algorithm across the 100 random trials for different numbers of  $N_{pilot}$ and  $N_{iter}$, represented as histograms. Compared with the variance reduction curve that is being maximized\footnote{Note that maximizing the variance reduction is equivalent to minimizing the estimator variance which was the approach implemented in this work (Equation \eqref{eq:bi_level_tuning})}, it can be seen in (a) that the EGO-optimized time steps noticeably improve with increasing $N_{pilot}$. In particular, for $N_{pilot}=10$, it can be seen that a relatively large percentage of optimized time steps are larger values that are sub-optimal in the sense that the variance reduction is low. For increasing  $N_{iter}$ in (b), the trend is less clear, indicating that  $N_{iter}=5$ seems to be sufficient and there are diminishing returns beyond this value. More detailed optimization results can be found in Appendix \ref{sec:appendix_ex1_opt_results} and Figure \ref{fig:1d_opt_results_vs_num_iteration}.  Note that while the variance reduction objective in Figure \ref{fig:1d_tuning_results} appears simple to optimize, this curve was generated using $N_{pilot}=10k$ and is particularly noisy for the smaller values of $N_{pilot}$ used for model tuning, making the model tuning optimization problem challenging even in one dimension. This challenge is expanded upon in Appendix \ref{sec:appendix_ex1_opt_noise} and Figure \ref{fig:1d_example_sweep_trials}. As noted in Section \ref{section:optimization_approach}, the online pilot sampling approach could be used rather than pilot  projection during model tuning optimization to yield more precise variance estimates, at the 
expense of adaptively evaluating additional pilot samples.

Figure \ref{fig:1d_tuning_overhead} presents the computational cost of the model tuning optimization problem for different numbers of  $N_{pilot}$ and  $N_{iter}$. Here, the model tuning overhead is represented as a percentage of the total budget, $100 \times \frac{C_{\mathcal{T}}}{C_{budget}}$, for different total budgets $C_{budget}=\{500,1000, 2000\}$. Since there is variability in the progression of the EGO algorithm, the overheads are shown as box plots illustrating the variation across the 100 random trials. Intuitively, this shows that the overhead incurred by model tuning prior to multifidelity uncertainty propagation is less significant with increasing total computational budget. So for $C_{budget}=500$, using $N_{pilot}=100$ or $N_{iter}=20$ depletes a large percentage of the budget (between $10-30\%$), while these percentages drop below $5\%$ for $C_{budget}=2000$. Table \ref{tab: tuning_results_1} summarizes the performance obtained through model tuning optimization along with the incurred overhead cost. Here, the $\%$ of random model tuning trials that returned hyperparameters that result in lower estimator variance than the \textbf{Hand Selected} solution are reported along with the average tuning cost (in terms of number of high-fidelity simulations). It can be seen that performance increases with increasing $N_{pilot}$, albeit with diminishing returns, while performance \textit{does not} increase with increasing $N_{iter}$, despite dedicating increased cost for model tuning. This result indicates that it is more worthwhile to consider larger values for $N_{pilot}$ rather than $N_{iter}$ for this one-dimensional tuning problem.

\begin{table}[h!]
  \small
  \centering
  \caption{Model tuning optimization results for example 1 for different numbers of EGO iterations ($N_{iter}$) and pilot samples ($N_{pilot}$).}
  \label{tab: tuning_results_1}
  \begin{tabular}{cccc}
    \hline \hline
    $N_{iter}$ & $N_{pilot}$ & $\%$ Improved Trials & Avg. Tuning Cost  \\
    \hline
    5  & 10  & 68 & 6.8 \\
    5  & 50  & 91 & 35.0 \\
    5  & 100 & 97 & 72.9 \\
    \hline
    5  & 25  & 84 & 17.1 \\
    10 & 25  & 82 & 29.1 \\
    20 & 25  & 81 & 44.6 \\
    \hline \hline
  \end{tabular}
\end{table}

\begin{figure}
    \centering
    \includegraphics[width=3in]{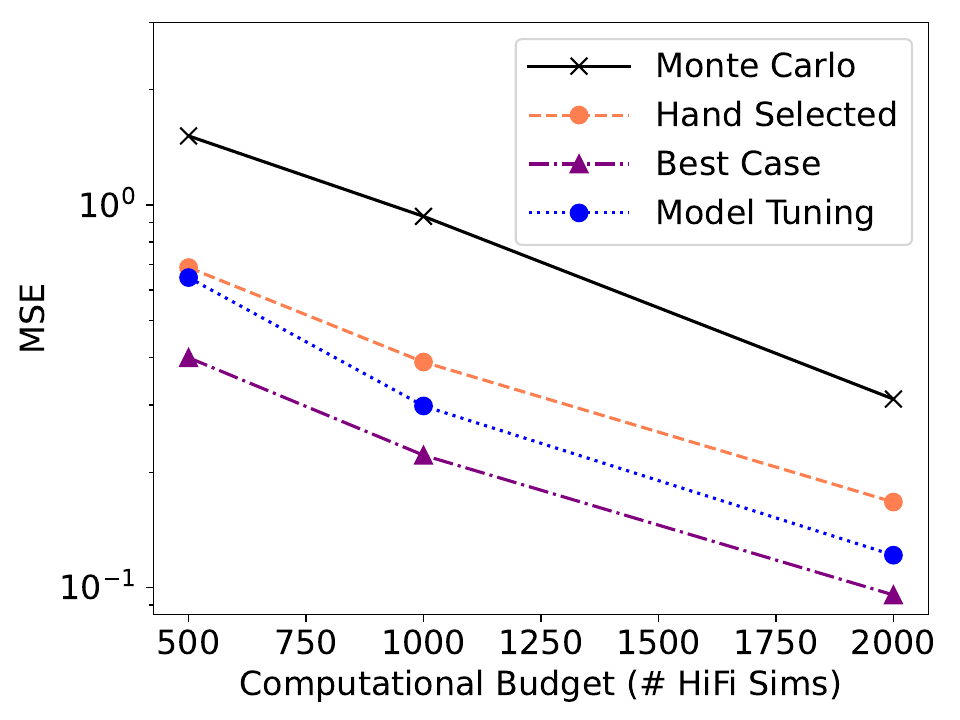}
    \caption{MSE versus budget for model tuning ($N_{pilot}=50$, $N_{iter}=5$) versus hand-tuned models, best-case ACV, and Monte Carlo solutions.}
    \label{fig:1d_mse_vs_budget}
\end{figure}

 The MSE of the estimators resulting from model tuning are compared with the other solutions for different computational budgets in Figure \ref{fig:1d_mse_vs_budget}. This model tuning result here depicts the particular case of $N_{pilot}=50$ and $N_{iter}=5$. It is clear that all three ACV estimators improve significantly upon the \textbf{Monte Carlo} solution, with the model tuning solution achieving performance between that of the \textbf{Hand Selected} and \textbf{Best Case} solutions across all budgets. However, the model tuning accuracy is comparable to the  \textbf{Hand Selected} solution for $C_{budget} = 500$ when the cost of the model tuning optimization relative to the total budget is higher.

\begin{figure}[htbp]
    \centering
    \includegraphics[width=\textwidth]{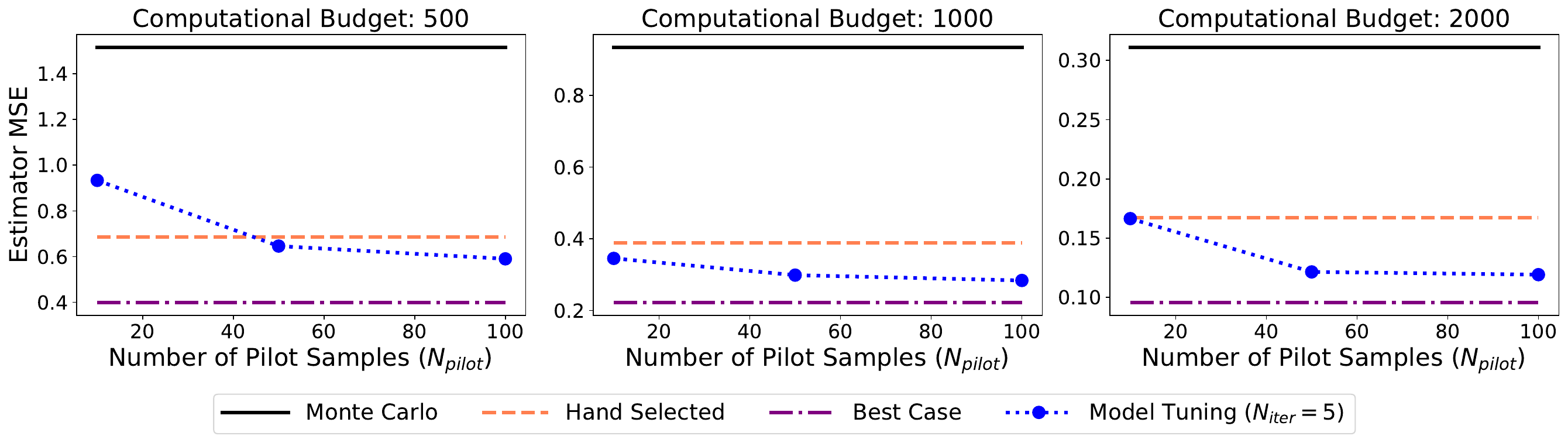}
    \caption{MSE versus number of pilot samples compared with the Monte Carlo, hand-selected, and best case estimators for different computational budgets.}
    \label{fig:1d_mse_vs_num_pilot}
\end{figure}

\begin{figure}[htbp]
    \centering
    \includegraphics[width=\textwidth]{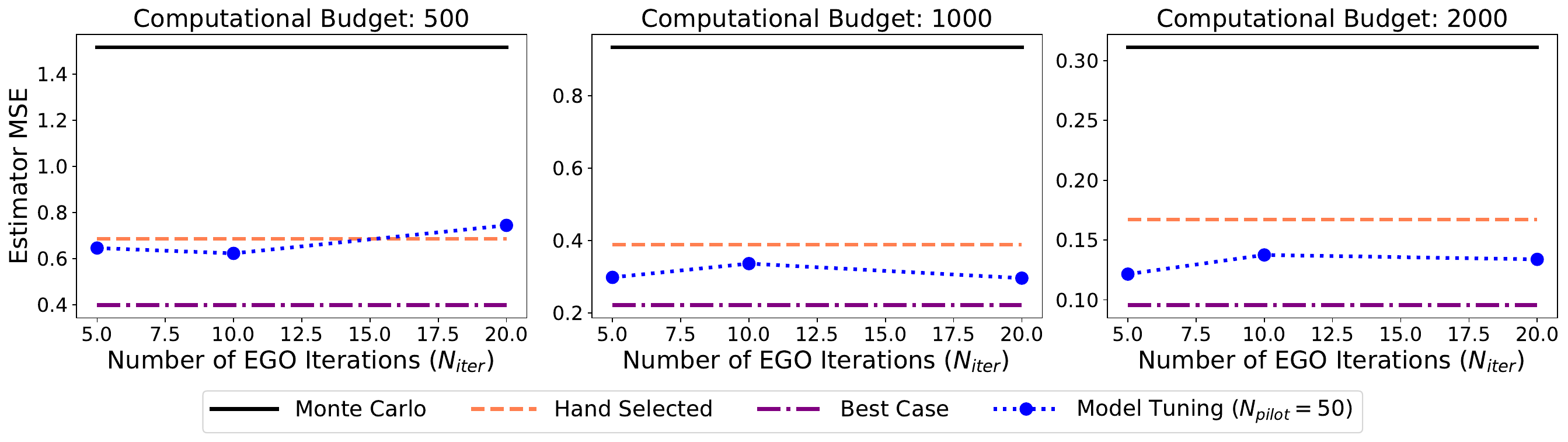}
    \caption{MSE versus number of EGO iterations compared with the Monte Carlo, hand-selected, and best case estimators for different computational budgets.}
    \label{fig:1d_mse_vs_num_iter}
\end{figure}

Finally, the MSE for the model tuning estimators is viewed as a function of $N_{pilot}$ in Figure \ref{fig:1d_mse_vs_num_pilot} and $N_{iter}$ in Figure \ref{fig:1d_mse_vs_num_iter} for different computational budgets. There is a clearer trend of improved performance for increasing values of $N_{pilot}$ than there is for increasing $N_{iter}$, reflecting that the overhead incurred by additional EGO iterations is not worth the negligible performance gains from slightly improved $\Delta t$ values. This observation is consistent with the optimized time steps reported for different  $N_{iter}$ in Figure \ref{fig:1d_tuning_results}. Overall, it is shown that improvement in MSE is obtained with model tuning versus the \textbf{Hand Selected} for a variety of parameter ($N_{iter}$ and $N_{pilot}$ ) settings, even for relatively modest computational budgets. However, in most cases, there is still a significant gap in performance with respect to the  \textbf{Best Case} solution that likely requires larger $C_{budget}$ to close. Since it was not computationally tractable to investigate $C_{budget}> 2000$ when multiple trials are required to calculate MSE, Example 2 assesses estimator performance in terms of predicted variance and is thus not constrained by $C_{budget}$.

 \subsection{Example 2 - Tuning of Multiple Hyperparameters}

Example 2 studies the tuning of time steps for both the Reduced Physics, $\Delta t_{RP}$,  and Coarse Time Step, $ \Delta t_{CST}$,  models through the solution of Equation \eqref{eq:bi_level_tuning}. Similar to Example 1, model tuning performance is evaluated for different numbers of pilot samples, $N_{pilot} = \{10, 50, 100\}$, and EGO iterations, $N_{iter}=\{10, 20, 30\}$\footnote{EGO uses an initial sample size $N_{init}=6$  to generate a surrogate model in two dimensions, so these values represent 4, 14, and 24 EGO iterations beyond the initial sample, respectively}. Performance was assessed in terms of estimator variance, rather than the more computationally-intensive MSE considered in Example 1, allowing for the consideration of additional QoIs (the landing latitude and angle of attack at 10km) and larger computational budgets. The variance reduction contours for each of the three QoIs can be seen in Figure \ref{fig:variance_reduction_contours} (b)-(d), showing the varied nature of the model tuning optimization problems (perhaps simpler for landing latitude relative to time of flight and angle of attack). A maximum computational budget of $C_{budget}=10000$ is considered to better understand the relative advantages of model tuning for more realistic analyses\footnote{The original Monte Carlo analysis for ADEPT used $C_{budget}=8000$ high-fidelity simulations \cite{dutta2019flight}}. Note that while the model tuning procedure is simply repeated for each QoI individually for the results presented next, the approach could be extended to find hyperparameters that perform optimally for the joint estimation of all QoIs simultaneously using multioutput ACVs \cite{doi:10.1137/23M1607994}.

\begin{figure} 
\centering 
\subfigure[]{\includegraphics[width=2.75in]{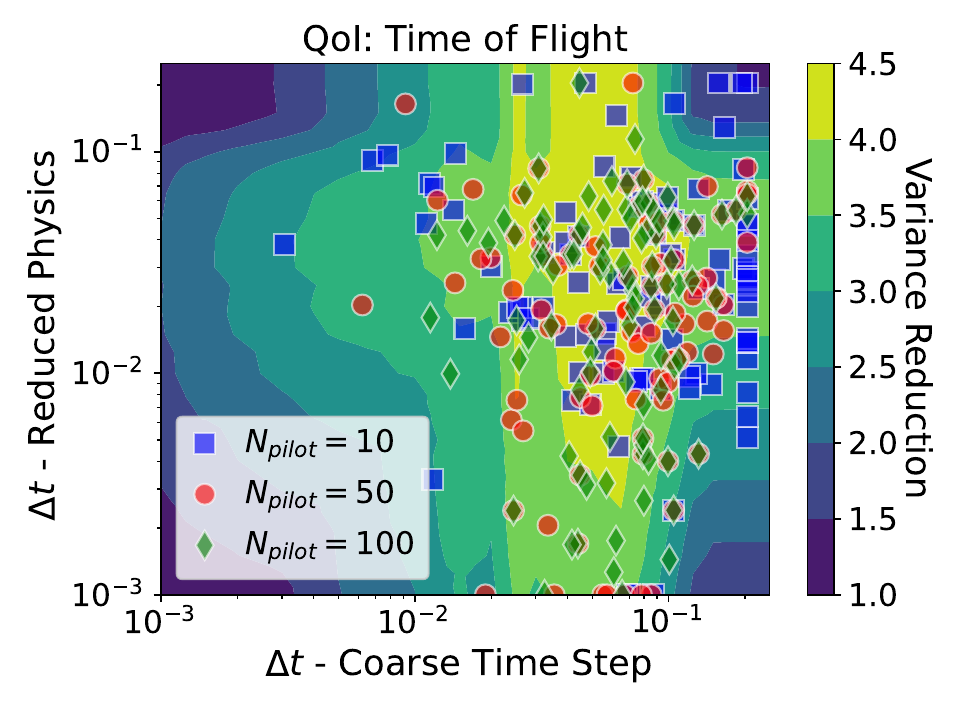}}
\subfigure[]{\includegraphics[width=2.75in]{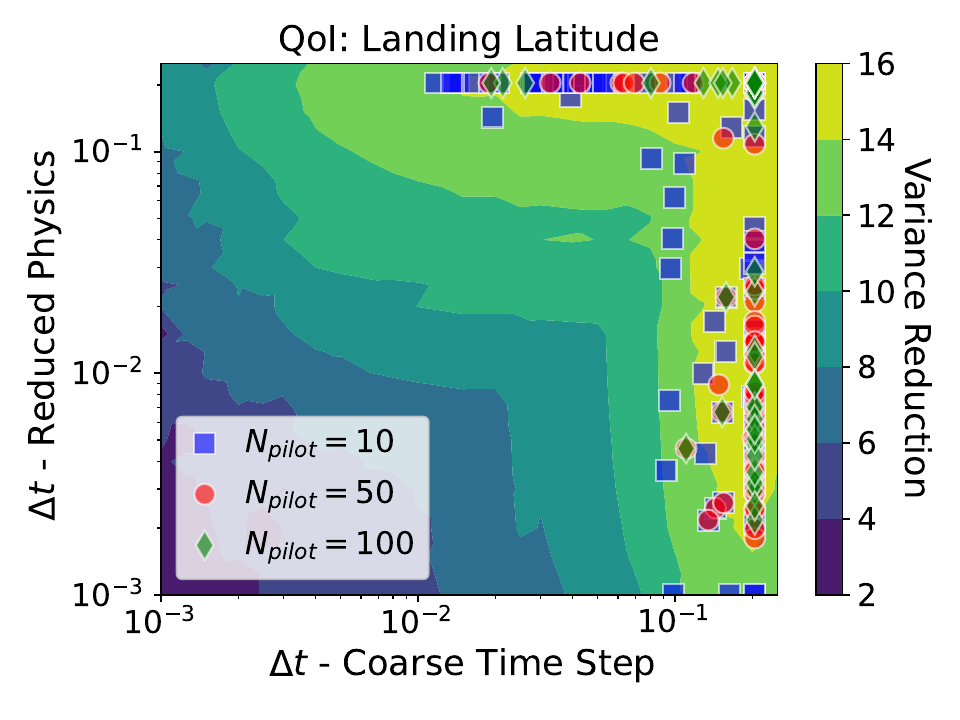}}
\subfigure[]{\includegraphics[width=2.75in]{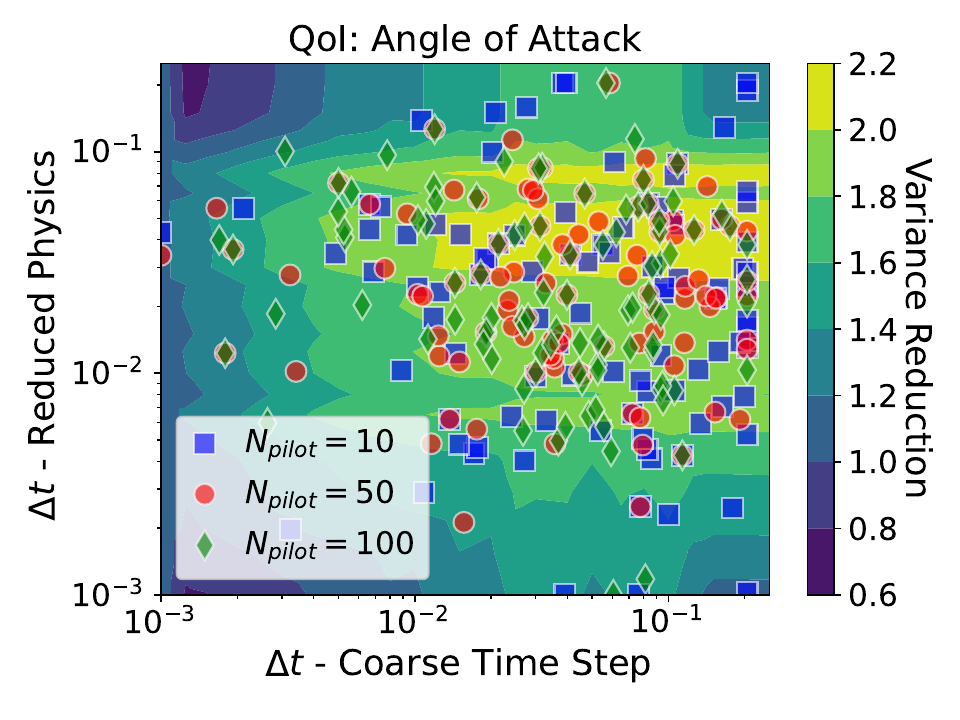}}
\caption{Model tuning optimization results across random trials for different numbers of pilot samples for a) time of flight, b) landing latitude, and c) angle of attack QoIs.} \label{fig:2d_opt_results}
\end{figure}

\begin{figure} 
\centering 
\subfigure[]{\includegraphics[width=2.75in]{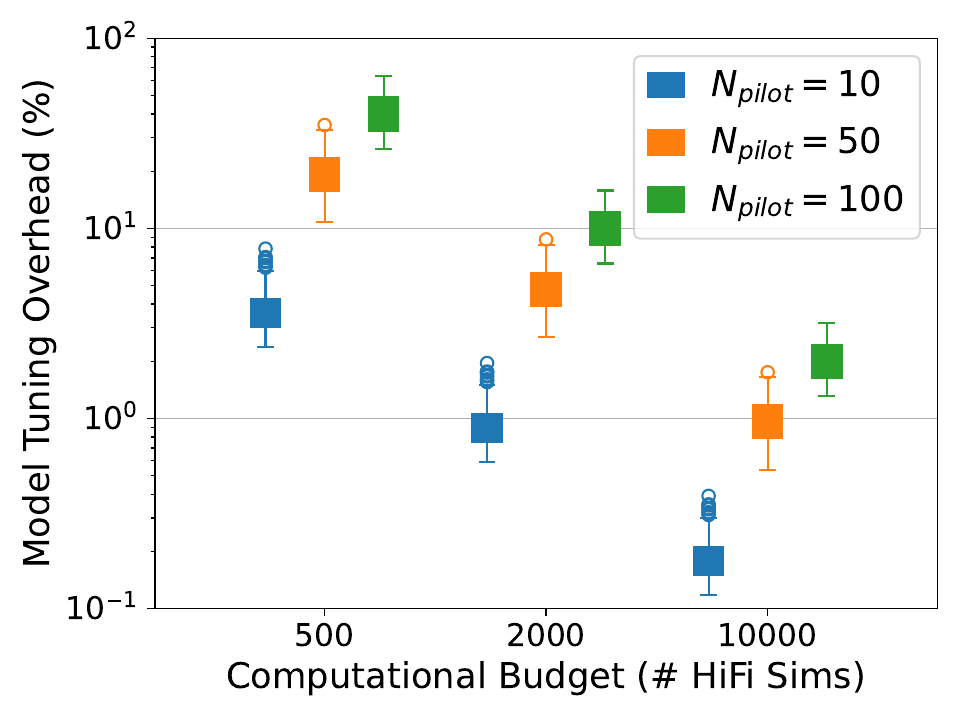}}
\subfigure[]{\includegraphics[width=2.75in]{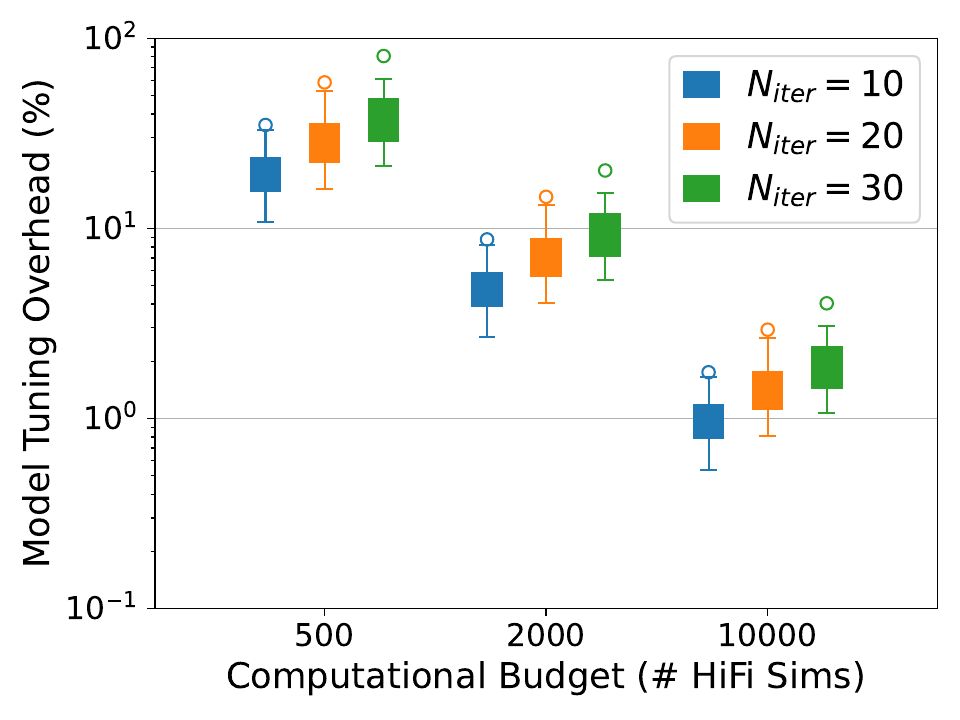}}
\caption{The computational overhead ($\%$) from solving the model tuning optimization problem in example 2 for different numbers of a) pilot samples and b) EGO iterations.}\label{fig:2d_tuning_overhead}
\end{figure}

The optimized time steps and the associated computational cost from model tuning optimization are shown in Figures \ref{fig:2d_opt_results} and \ref{fig:2d_tuning_overhead}, respectively. Figure \ref{fig:2d_opt_results} shows the optimized time steps across 100 random trials superimposed on the variance reduction contours for each of the three QoIs. Generally speaking, the time steps become more closely clustered in the region of optimality for increasing number of pilot samples. A larger amount of variation in optimized time steps is observed for the angle of attack QoI, potentially indicating more noise and a more challenging optimization problem relative to the other QoIs. The optimized time steps for the landing latitude are clustered around the boundary of the optimization domain, indicating that the estimator variance is relatively constant for large time step values of either model. The computational cost of model tuning is shown in Figure  \ref{fig:2d_tuning_overhead}, again reported as a percentage of the total budget, $100 \times \frac{C_{\mathcal{T}}}{C_{budget}}$. It can be seen that when $C_{budget}=10000$, even larger settings of $N_{pilot}$ and $N_{iter}$ (and hence, improved estimates of $\Delta t^*_{RP}$ and $\Delta t^*_{CST}$) incur a relatively negligible overhead.

The performance of the model tuning solutions was assessed with respect to the other baseline approaches in terms of variance reduction with respect to the \textbf{Monte Carlo} solution (Equation \ref{eq:variance_reduction}). In order to reduce the computation time needed to estimate the variance reduction, an interpolator, $\tilde{\mathbb{V}}$, was constructed to approximate $\Var{\acvest}$
\begin{equation}\label{eq:variance_interpolant}
	\Var{\acvest} \approx \tilde{\mathbb{V}}(\Delta t_{RP}, \Delta t_{CST}).
\end{equation}
Here,  $\tilde{\mathbb{V}}$ is constructed from estimator variance data estimated on a  $25 \times 25$ grid of equally (log) spaced time steps using \textit{offline pilot} solution mode with $N_{pilot}=10k$ (i.e., the same data used to produce the contour plots in Figure \ref{fig:variance_reduction_contours}). For the \textbf{Hand Selected} solution, $\tilde{\mathbb{V}}$ was evaluated with $\Delta t_{RP}=0.001$ and $\Delta t_{CST}=0.1$ while the \textbf{Best Case} solutions used different optimal time step settings for the different QoIs considered (see Figure \ref{fig:variance_reduction_contours}). For the model tuning solution, the estimated variance from the interpolator was scaled to account for the computational overhead of solving Equation \eqref{eq:bi_level_tuning},
\begin{equation}\label{eq:model_tuning_var}
	\Var{\acvest} \approx \tilde{\mathbb{V}}(\Delta t^*_{RP}, \Delta t^*_{CST}) \times \frac{C_{budget}}{C_{budget}-C_{\mathcal{T}}},
\end{equation}
where $\Delta t^*_{RP}$ and  $\Delta t^*_{CST}$ are the optimized time steps found using the EGO algorithm. This simple approach is possible since  $\Var{\acvest} \propto \frac{1}{C_{budget}}$. When the model tuning overhead is high, then it can be seen than Equation \eqref{eq:model_tuning_var} increases the estimated variance accordingly to account for the decreased proportion of the computational budget available for ACV variance reduction.

\begin{figure} 
\centering 
\subfigure[]{\includegraphics[width=2.75in]{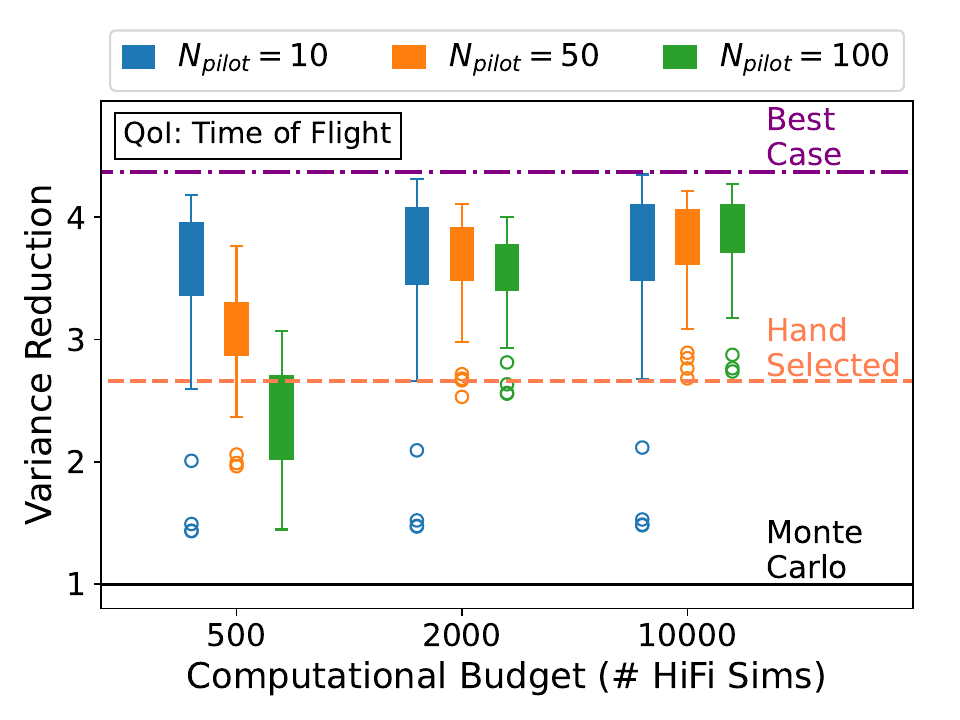}}
\subfigure[]{\includegraphics[width=2.75in]{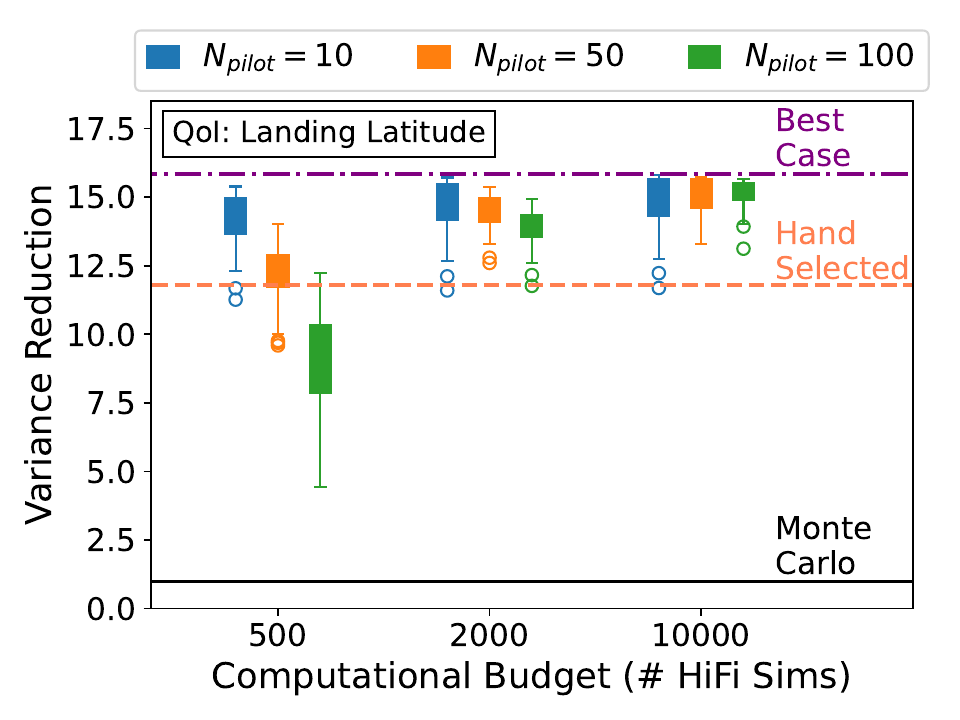}}
\subfigure[]{\includegraphics[width=2.75in]{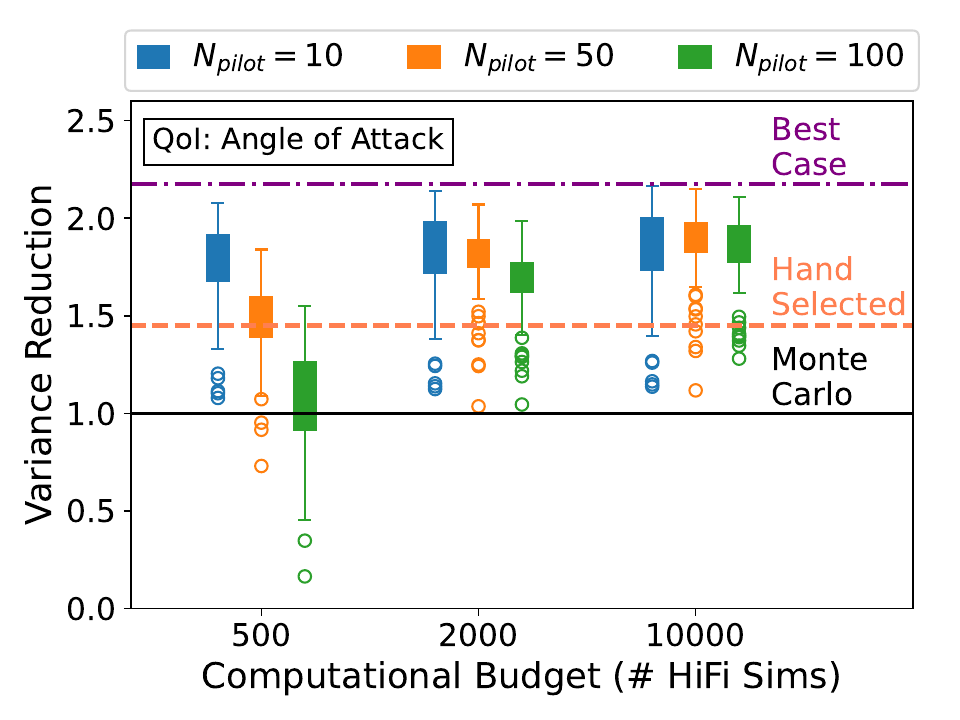}}
\caption{Model tuning performance for different numbers of pilot samples and computational budgets for a) time of flight, b) landing latitude, and c) angle of attack QoIs.}\label{fig:2d_var_red_vs_num_pilot}
\end{figure}

\begin{figure} 
\centering 
\subfigure[]{\includegraphics[width=2.75in]{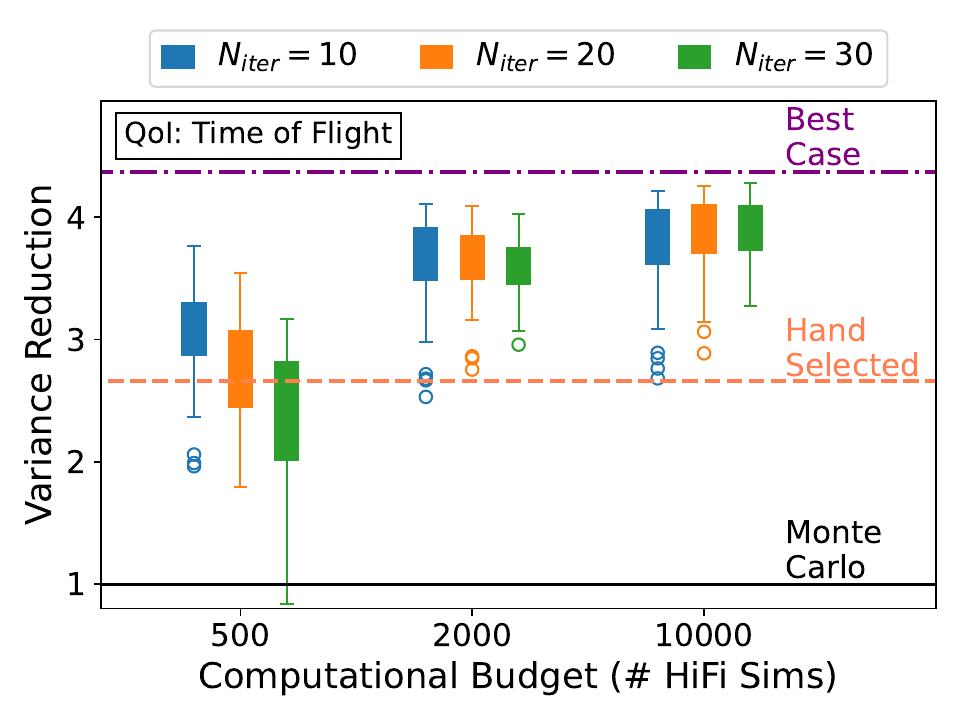}}
\subfigure[]{\includegraphics[width=2.75in]{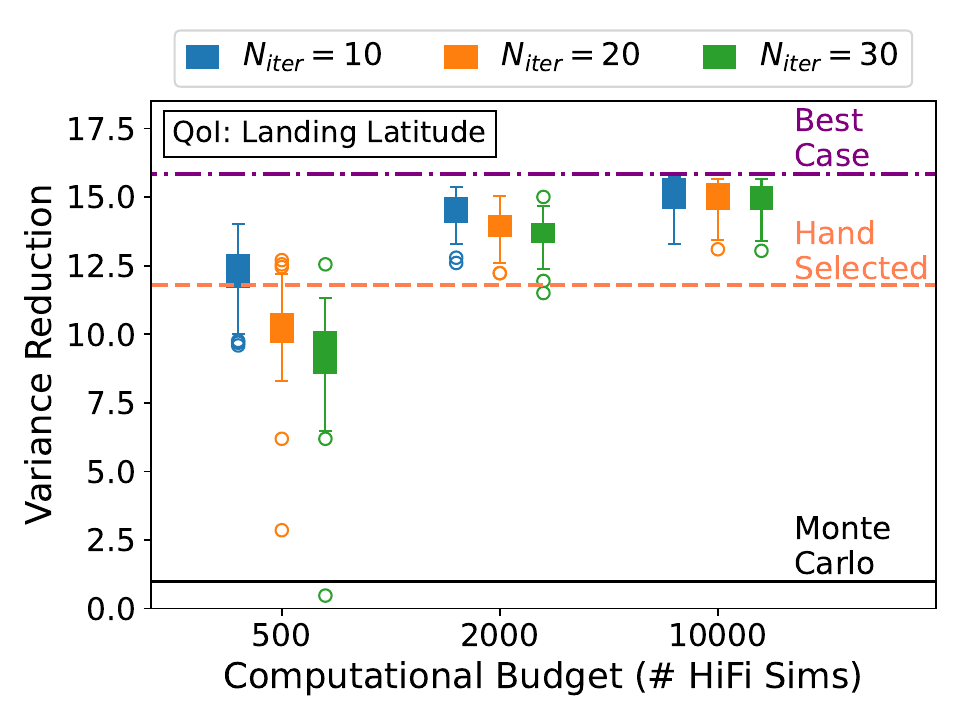}}
\subfigure[]{\includegraphics[width=2.75in]{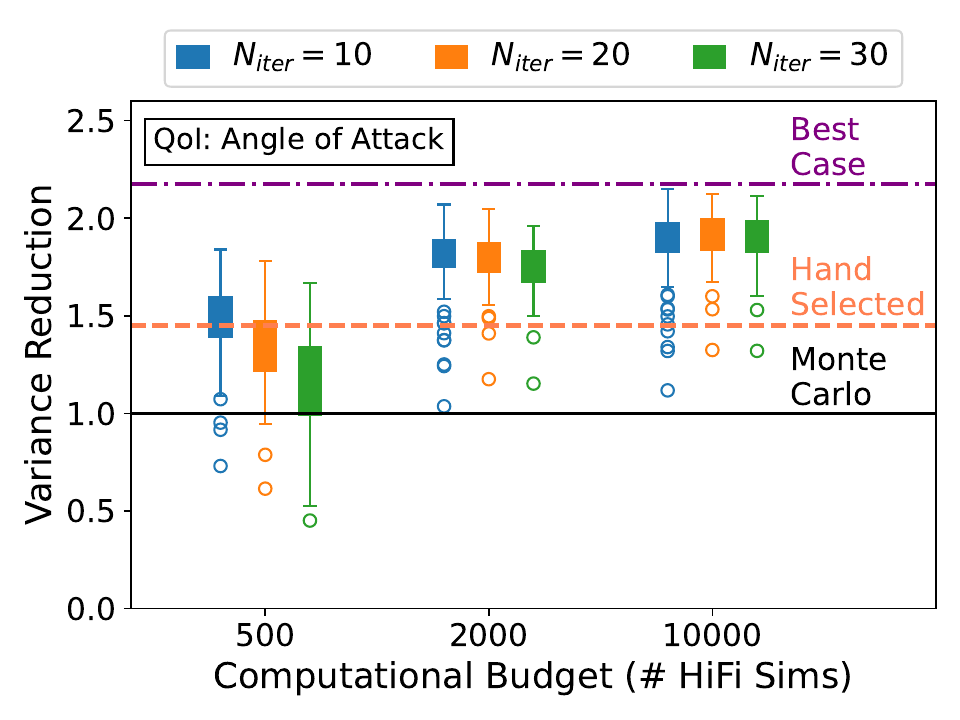}}
\caption{Model tuning performance for different numbers of EGO iterations and computational budgets for a) time of flight, b) landing latitude, and c) angle of attack QoIs.}\label{fig:2d_var_red_vs_num_iter}
\end{figure}

The variance reduction achieved by ACV estimators with each approach is compared for varying numbers of pilot samples and EGO iterations in Figures \ref{fig:2d_var_red_vs_num_pilot} and \ref{fig:2d_var_red_vs_num_iter}, respectively.  Intuitively, as the computational budget increases from $500$ to $10000$, the relative performance of the model tuning estimators increases for all QoIs as the model tuning overhead becomes a more negligible portion of the overall budget. Along these lines, for the smallest budget, $C_{budget} = 500$, it is shown that seeking more accurate model tuning solutions with larger $N_{pilot}$ and $N_{iter}$ results in lower performance. For example, the variance reduction resulting from $N_{pilot}=100$ and $N_{iter}=30$ is generally lower than for estimators using the simpler \textbf{Hand Selected} approach, and in some cases (e.g., for angle of attack) shows performance worse than a standard Monte Carlo approach. On the other hand, as the total computational budget increases, $C_{budget} = 10000$, it becomes worthwhile to allocate more resources to the model tuning optimization problem with larger values of $N_{pilot}$ and $N_{iter}$. It can be seen that variance reductions for these cases tend to approach and eventually surpass the performance provided by lower $N_{pilot}$ and $N_{iter}$ settings as the budget increases. Overall, model tuning is shown to be largely beneficial, provided that the computational budget is relatively large, and that the performance of the resulting estimators eventually approaches the best case performance possible with increasingly large budget.

 \subsubsection*{Model Tuning with Parametric ACV Search}
 
 \begin{figure} 
\centering 
\subfigure[]{\includegraphics[width=2.9in]{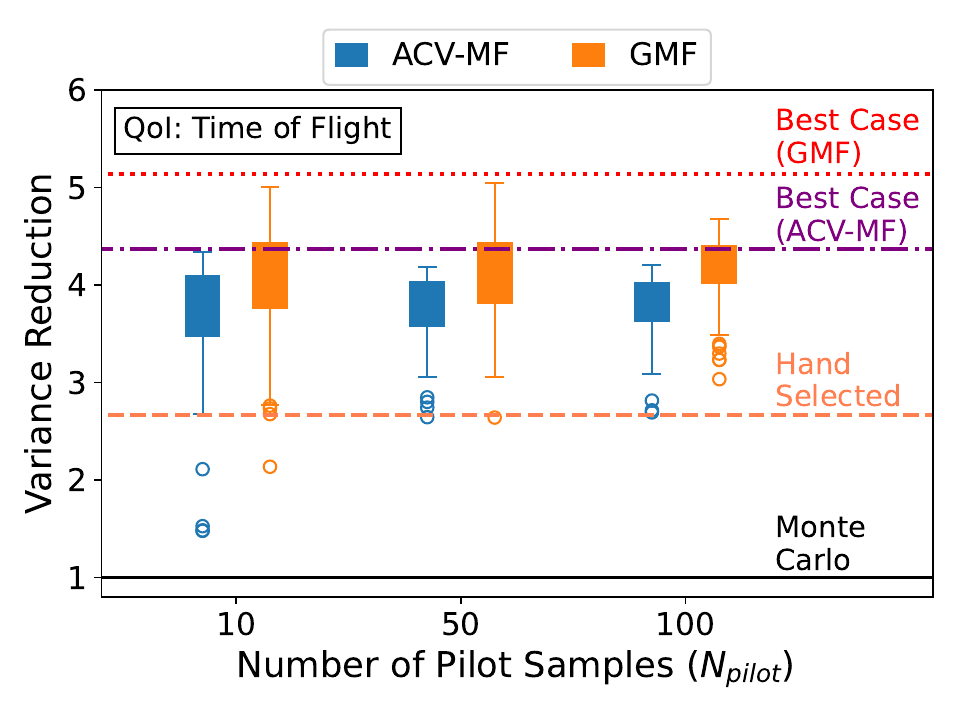}}
\subfigure[]{\includegraphics[width=2.9in]{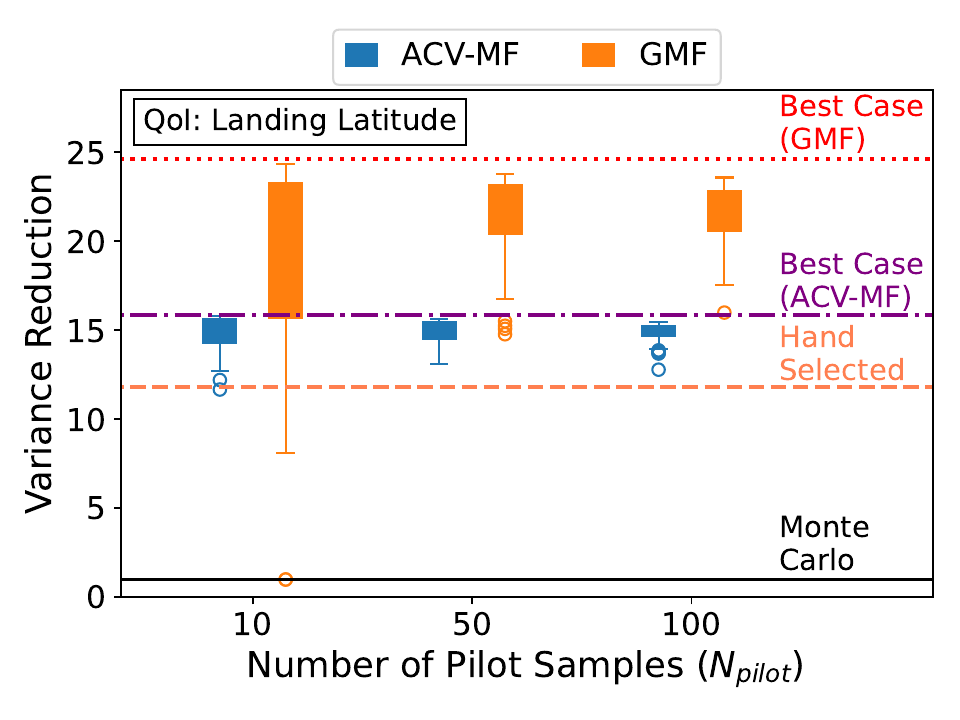}}
\subfigure[]{\includegraphics[width=2.9in]{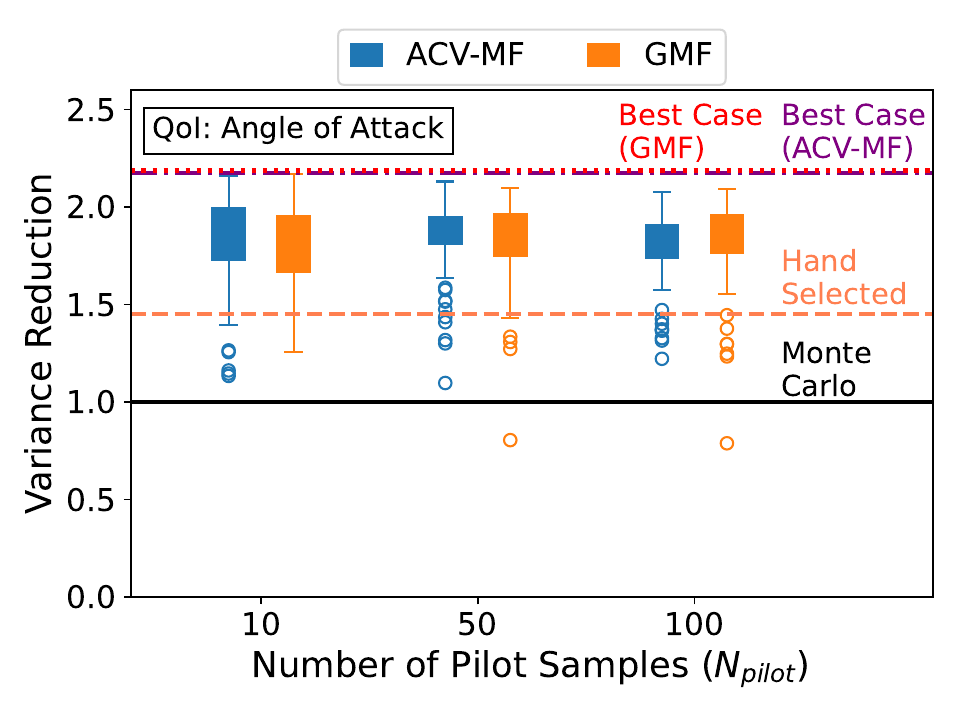}}
\caption{Model tuning performance when combined with parametric ACV search for different numbers of pilot samples for a) time of flight, b) landing latitude, and c) angle of attack QoIs.} \label{fig:model_tuning_with_parametric_acv}
\end{figure}

This section shows results that combine model tuning with a parametric ACV search, described in Section \ref{sec: model_tuning_problem}, demonstrating the performance gains that are possible from the added flexibility in the tuning process. Specifically, the generalized multifidelity estimator (GMF) \cite{BOMARITO2022110882}, which searches over sixteen recursion trees that are admissible for the case of four models, is considered for the inner loop optimization in Equation \eqref{eq:bi_level_tuning}. Note that the ACV-MF estimator used in the previous examples is one such possible choice. The approach was performed for all three QoIs with varying numbers of pilot samples for a fixed computational budget of 5000 and ten EGO iterations. One hundred random trials were performed once again for each case. 
 
 The results from this study are shown in Figure \ref{fig:model_tuning_with_parametric_acv} for the a) time of flight, b) landing latitude, and c) angle of attack QoIs, comparing the performance of incorporating the parametric ACV search (GMF) with the  ACV-MF results from the previous example. The ``best case" variance reduction for GMF is indicated by an additional  horizontal line to show the maximum possible performance increase possible relative to ACV-MF.  It can be see that the performance gains from incorporating the parametric ACV search are highly dependent on the QoI. The variance reduction for the landing latitude improved substantially, with the best case performance increasing by over $50\%$ and nearly all of the random trials achieving higher variance reduction relative to the fixed ACV-MF estimator. On the other hand, the variance reduction for the angle of attack QoI remained largely unchanged, indicating that the optimal recursion tree found with the parametric ACV search was that of ACV-MF. While parametric ACV will in general provide improved performance, the computational cost incurred during model tuning can become significant if the number of models in the ensemble is large.

\section{Discussion and Conclusions}

The practicality of automated model tuning for multifidelity uncertainty propagation  was explored in the context of an EDL application. When considered an offline cost, expending computational resources on tuning low-fidelity models to perform better with multifidelity estimators is a largely worthwhile exercise. However, for online analyses, such as those required for onboard trajectory simulation, the cost vs. precision tradeoff posed by model tuning optimization must be carefully considered. 

To explore this tradeoff, this study focused on implementing online multifidelity estimators using model tuning for a budget-constrained trajectory simulation problem. The time steps of two low-fidelity flight mechanics models were considered as tunable hyperparameters. A bi-level optimization strategy was employed to solve the model tuning problem, with the EGO algorithm on the outer loop and the  ACV sample allocation problem on the inner loop, both implemented using Sandia National Laboratories Dakota software. The performance of using low-fidelity models that were tuned on-the-fly as part of an online multifidelity uncertainty propagation analysis was assessed in terms of their variance and mean squared error with respect to three baseline approaches: 1) standard Monte Carlo, 2) ACVs with hand-tuned low-fidelity models, and 3) ACVs with globally optimal low-fidelity models, representing the best-case performance. The comparison was carried out for varying values of parameters that control the accuracy and cost of the model tuning optimization problem, namely the number of EGO iterations and the number of pilot samples used in each iteration to solve the inner loop ACV problem. Three trajectory simulation quantities of interest were considered across a range of total computational budgets.

The results confirmed the intuition that the feasibility and usefulness of automated model tuning, when considering the overhead incurred, increases with the overall computational budget afforded for multifidelity uncertainty propagation with ACV estimators. For the particular EDL application studied, it was found  incorporating model tuning largely outperforms ACV estimators with hand-tuned models, even when the total computational budget is relatively low. As the computational budget grows, estimators using model tuning approach the best-case ACV performance possible when optimal hyperparameters and model statistics are known \textit{a priori}. Moreover, it was shown that incorporating a parametric ACV search over permissible model recursion graphs within the model tuning optimization could increase the potential variance reduction significantly.  It is also worth noting that the viability of hand-tuning of hyperparameters is questionable in the case of on-board trajectory simulation, where environments and trajectories might be changing significantly from nominal values.  In these cases, despite the small computational budget that may be afforded, on-the-fly model tuning is likely the best option.

While this study sheds light on the utility of model tuning in a realistic setting, there remain open questions and avenues for improving the viability of such approaches. For example, the findings reported here with only two tunable low-fidelity models likely do not reflect the challenges that would be encountered in higher-dimensional model tuning problems. When several tunable models are present, the resulting optimization problem gets more complex, but the potential performance improvements likely grow significantly as hand-tuning becomes infeasible. In terms of future avenues for algorithmic improvements, employing advanced Bayesian optimization algorithms for solving model tuning such as those that are cost-aware \cite{guinet2020paretoefficientacquisitionfunctionscostaware} or more robust to noise \cite{doi:10.1287/ijoc.1080.0314} would help navigate the tradeoff of estimator cost versus precision with more rigor. Alternatively, a more dynamic assessment of diminishing returns from model tuning could be used to terminate the model tuning optimization problem, rather than the fixed convergence criteria considered here based on maximum number of iterations. Finally, this work adopted online pilot (iterated ACV) \cite{dakota2024} as a simple method to handle unknown model statistics for multifidelity estimation, but more advanced adaptive approaches from recent work \cite{xu2022banditlearningapproachmultifidelityapproximation, coons2025bayesiancovarianceuncertaintyadaptive,dixon2025optimallybalancingexplorationexploitation} could further improve estimator performance.

\section*{Acknowledgements}
\noindent James Warner and Geoffrey Bomarito carried out this  work under the \textit{Entry Systems Modeling Project}, which was funded by NASA's Game Changing Development Program within the Space Technology Mission Directorate.

\noindent Sandia National Laboratories is a multi-mission laboratory managed and operated by National Technology \& Engineering Solutions of Sandia, LLC (NTESS), a wholly owned subsidiary of Honeywell International Inc., for the U.S. Department of Energy’s National Nuclear Security Administration (DOE/NNSA) under contract DE-NA0003525. This written work is authored by an employee of NTESS. The employee, not NTESS, owns the right, title and interest in and to the written work and is responsible for its contents. Any subjective views or opinions that might be expressed in the written work do not necessarily represent the views of the U.S. Government. The publisher acknowledges that the U.S. Government retains a non-exclusive, paid-up, irrevocable, world-wide license to publish or reproduce the published form of this written work or allow others to do so, for U.S. Government purposes. The DOE will provide public access to results of federally sponsored research in accordance with the DOE Public Access Plan.

\bibliography{main}

\begin{appendices}
    \section{ACV Estimator Variance}
The ACV estimator is given by
\begin{equation*}
    \acvest\left(\hyper,\avec,\zvec\right)  = \mcest(\zvec) + \avec^\mathrm{T} \diffvec(\hyper,\zvec)\\
\end{equation*}
The variance of this estimator is obtained through the variance of scaled random variables and the variance of sums of random variables:
\begin{equation*}
\Var{\acvest\left(\hyper,\zvec \right)}  = \Var{\mcest(\zvec)}  + 
                                         \avec^\mathrm{T} \covop{\diffvec(\hyper,\zvec)}{\diffvec(\hyper,\zvec)} \avec + 2\avec^\mathrm{T} \covop{\diffvec(\hyper,\zvec)}{\mcest(\zvec)}
\end{equation*}
Plugging in the optimal $\avec^{opt}$ from Equation \eqref{eq: optimal_alpha} then gives
\begin{equation*}
     \Var{\acvest\left(\hyper,\zvec \right)}  = \Var{\mcest(\zvec)}  - 
                                         \covop{\diffvec(\hyper,\zvec)}{\mcest(\zvec)}^\mathrm{T} \covop{\diffvec(\hyper,\zvec)}{\diffvec(\hyper,\zvec)}^{-1} 
                                         \covop{\diffvec(\hyper,\zvec)}{\mcest(\zvec)} \\
\end{equation*}
As shown in detail in \cite{BOMARITO2022110882}, the covariance of MC estimators, sample independence, covariance of sums, and the use of model statistics can be used to derive that: 
\begin{equation*}
    \covop{\diffvec(\hyper,\zvec)}{\diffvec(\hyper,\zvec)} = \mathbf{G}(\zvec) \circ \covmat(\hyper)
\end{equation*}
\begin{equation*}
    \covop{\diffvec(\hyper,\zvec)}{\mcest(\zvec)} = \mathbf{g}(\zvec) \circ \covvec(\hyper)
\end{equation*}
with
\begin{equation*}
G_{ij} = \frac{n(\sset{i}{1} \cap \sset{j}{1})}{n(\sset{i}{1})n(\sset{j}{1})} - \frac{n(\sset{i}{1} \cap \sset{j}{2})}{n(\sset{i}{1})n(\sset{j}{2})} - \frac{n(\sset{i}{2} \cap \sset{j}{1})}{n(\sset{i}{2})n(\sset{j}{1})} + \frac{n(\sset{i}{2} \cap \sset{j}{2})}{n(\sset{i}{2})n(\sset{j}{2})}
\end{equation*}
\begin{equation*}
C_{ij} = \covop{Q_i}{Q_j}
\end{equation*}
\begin{equation*}
 g_{ij} = \frac{n(\sset{i}{1} \cap \sset{}{})}{n(\sset{i}{1})n(\sset{}{})} - \frac{n(\sset{i}{2} \cap \sset{}{})}{n(\sset{i}{2})n(\sset{}{})}
\end{equation*}
\begin{equation*}
c_{i} = \covop{Q_i}{Q}
\end{equation*}
where $n()$ is the cardinality operator. The variance expression can be simplified further by making use of normalized values of both the cardinality and covariance. With that intent, define the operator $r() = \frac{n()}{n(\sset{}{})}$ as the normalized cardinality (also commonly called the oversampling ratio).  Thus
\begin{equation*}
    \begin{split}
    G_{ij} &= \frac{1}{n(\sset{}{})} \left[ \frac{r(\sset{i}{1} \cap \sset{j}{1})}{r(\sset{i}{1})r(\sset{j}{1})} - \frac{r(\sset{i}{1} \cap \sset{j}{2})}{r(\sset{i}{1})r(\sset{j}{2})} - \frac{r(\sset{i}{2} \cap \sset{j}{1})}{r(\sset{i}{2})r(\sset{j}{1})} + \frac{r(\sset{i}{2} \cap \sset{j}{2})}{r(\sset{i}{2})r(\sset{j}{2})} \right] \\
    &= \frac{1}{n(\sset{}{})} F_{ij}\\
    \mathbf{G} &= \frac{1}{N} \Fmat
    \end{split}
\end{equation*}
\begin{equation*}
    \begin{split}
    g_{ij} &= \frac{1}{n(\sset{}{})} \left[ \frac{r(\sset{i}{1} \cap \sset{}{})}{r(\sset{i}{1})} - \frac{r(\sset{i}{2} \cap \sset{}{})}{r(\sset{i}{2})} \right] \\
    &= \frac{1}{n(\sset{}{})} f_{ij}\\
    \mathbf{g} &= \frac{1}{N} \fvec
    \end{split}
\end{equation*}
Similarly, the covariance terms can be written in terms of correlation coefficients:
\begin{equation*}
    \begin{split}
    C_{ij} &= \sqrt{\Var{Q_i}\Var{Q_j}} \corrop{Q_i}{Q_j}\\
    &= \sigma_i \sigma_j \rho_{ij} \\
    \covmat &= \mathrm{diag}(\underline{\sigma}) \mathbf{P} \mathrm{diag}(\underline{\sigma}) 
    \end{split}
\end{equation*}
\begin{equation*}
    \begin{split}
    c_{i} &= \sqrt{\Var{Q_i}} \corrop{Q_i}{Q} \sqrt{\Var{Q}}\\
    &= \sigma_i \rho_{i} \sqrt{\Var{Q}} \\
    \covvec &= (\underline{\sigma} \circ \mathbf{\rho}) \sqrt{\Var{Q}} 
    \end{split}
\end{equation*}
where the standard deviations of the low-fidelity models are denoted $\underline{\sigma} = [\sigma_1, \dots, \sigma_M]^\mathrm{T}$.

The covariance terms are written in terms of these normalized values as
\begin{equation*}
    \covop{\diffvec(\hyper,\zvec)}{\diffvec(\hyper,\zvec)} = \frac{1}{N} \Fmat(\zvec) \circ \mathrm{diag}(\underline{\sigma}) \mathbf{P}(\hyper) \mathrm{diag}(\underline{\sigma}) 
\end{equation*}
\begin{equation*}
    \covop{\diffvec(\hyper,\zvec)}{\mcest(\zvec)} = \frac{\sqrt{\Var{Q}}}{N} \fvec(\zvec) \circ \underline{\sigma} \circ \mathbf{\rho}(\hyper) 
\end{equation*}
Substituting these into the equation for ACV estimator variance gives
\begin{equation*}
    \begin{split}
    \Var{\acvest\left(\hyper,\zvec \right)}  &= \Var{\mcest(\zvec)}  - 
        \left[ \frac{\sqrt{\Var{Q}}}{N} \fvec(\zvec) \circ \underline{\sigma} \circ \mathbf{\rho}(\hyper)\right]^\mathrm{T} 
        \left[\frac{1}{N} \Fmat(\zvec) \circ \mathrm{diag}(\underline{\sigma}) \mathbf{P}(\hyper) \mathrm{diag}(\underline{\sigma}) \right]^{-1} 
        \left[ \frac{\sqrt{\Var{Q}}}{N} \fvec(\zvec) \circ \underline{\sigma} \circ \mathbf{\rho}(\hyper)\right] \\
        &= \frac{\Var{Q}}{N}  - \frac{\Var{Q}}{N}
        \left[ \fvec(\zvec) \circ \underline{\sigma} \circ \mathbf{\rho}(\hyper)\right]^\mathrm{T} 
        \left[\Fmat(\zvec) \circ \mathrm{diag}(\underline{\sigma}) \mathbf{P}(\hyper) \mathrm{diag}(\underline{\sigma}) \right]^{-1} 
        \left[ \fvec(\zvec) \circ \underline{\sigma} \circ \mathbf{\rho}(\hyper)\right] \\
        &= \frac{\Var{Q}}{N}  - \frac{\Var{Q}}{N}
        \left[ \fvec(\zvec) \circ \mathbf{\rho}(\hyper)\right]^\mathrm{T} \circ \underline{\sigma}^\mathrm{T}
        \left[\mathrm{diag}(\underline{\sigma}) \left(\Fmat(\zvec) \circ  \mathbf{P}(\hyper)\right) \mathrm{diag}(\underline{\sigma}) \right]^{-1} 
        \underline{\sigma} \circ \left[ \fvec(\zvec) \circ  \mathbf{\rho}(\hyper)\right] \\
        &= \frac{\Var{Q}}{N}  - \frac{\Var{Q}}{N}
        \left[ \fvec(\zvec) \circ \mathbf{\rho}(\hyper)\right]^\mathrm{T}
        \left[\Fmat(\zvec) \circ  \mathbf{P}(\hyper)\right]^{-1} 
        \left[ \fvec(\zvec) \circ  \mathbf{\rho}(\hyper)\right] \\
        &= \Var{Q}\left( \frac{1}{N}  - \frac{1}{N}
        \left[ \fvec(\zvec) \circ \mathbf{\rho}(\hyper)\right]^\mathrm{T}
        \left[\Fmat(\zvec) \circ  \mathbf{P}(\hyper)\right]^{-1} 
        \left[ \fvec(\zvec) \circ  \mathbf{\rho}(\hyper)\right] \right)\\
    \end{split}
\end{equation*}

\section{Example 1: Detailed Model Tuning Optimization Results}\label{sec:appendix_ex1_opt_results}

\begin{figure} 
\centering 
\subfigure[]{\includegraphics[width=2.95in]{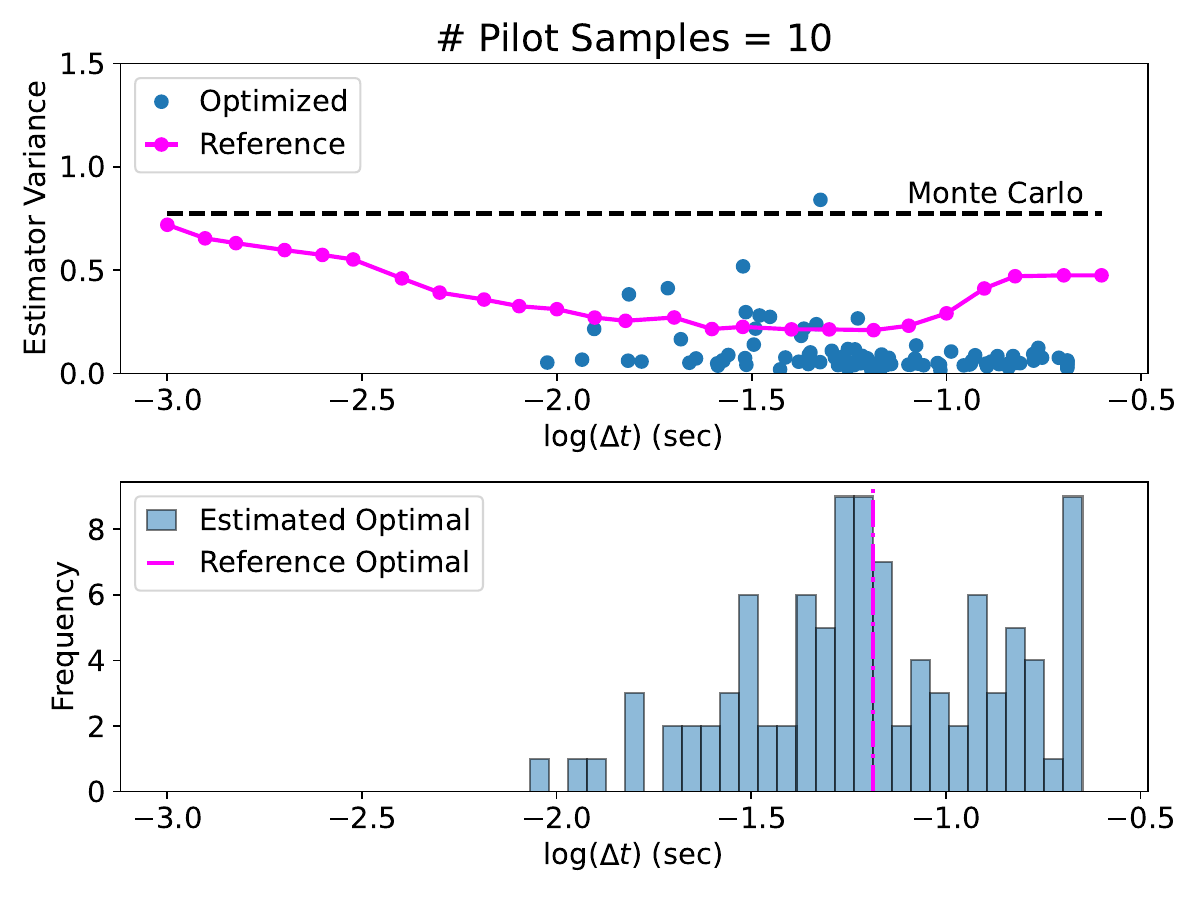}}
\subfigure[]{\includegraphics[width=2.95in]{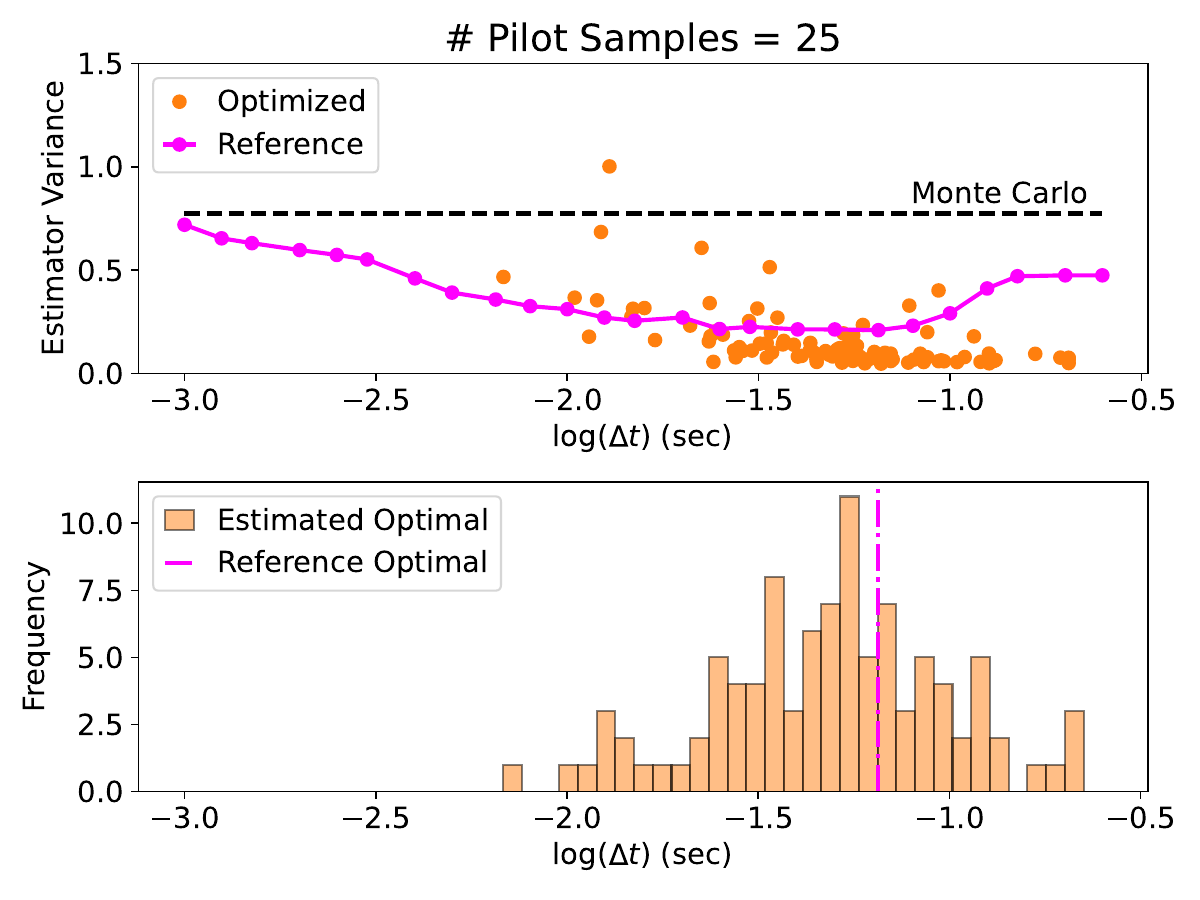}}
\subfigure[]{\includegraphics[width=2.95in]{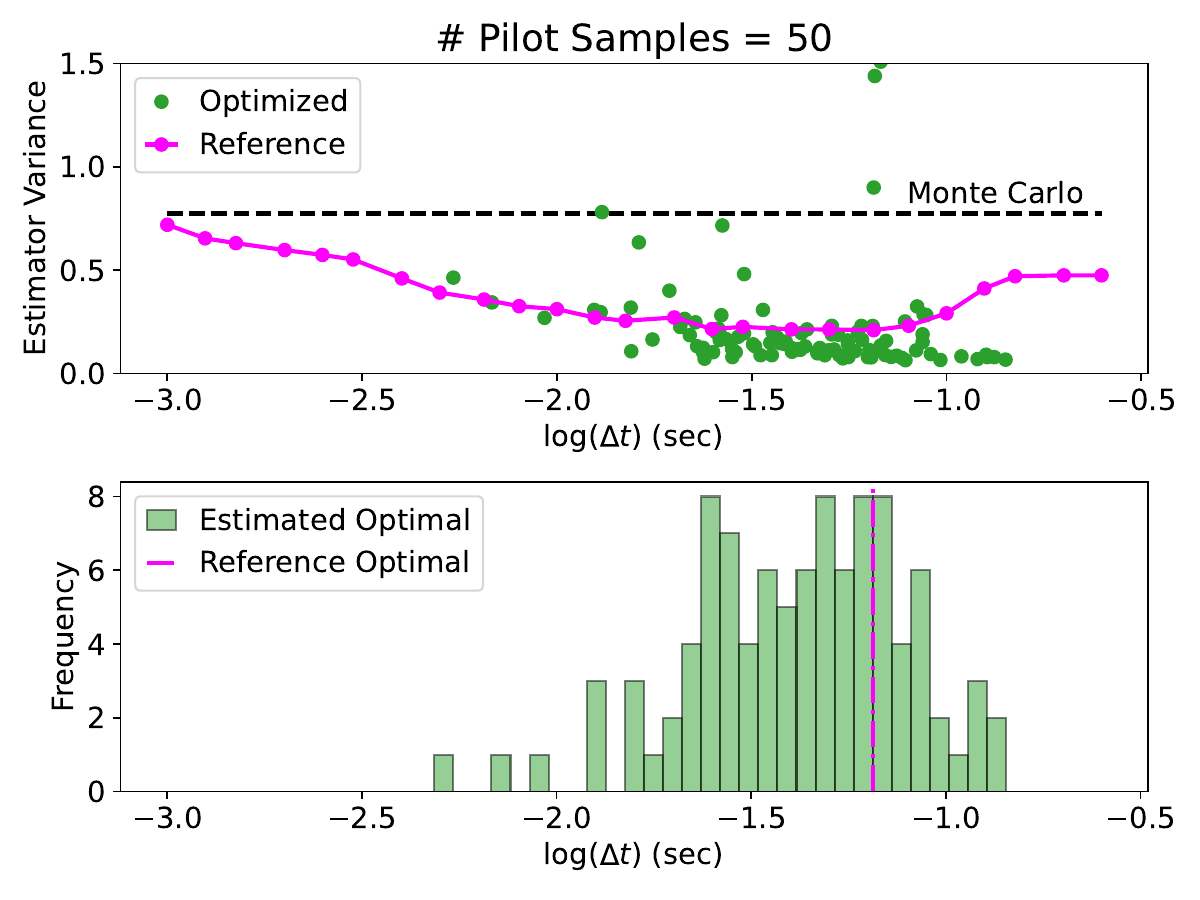}}
\subfigure[]{\includegraphics[width=2.95in]{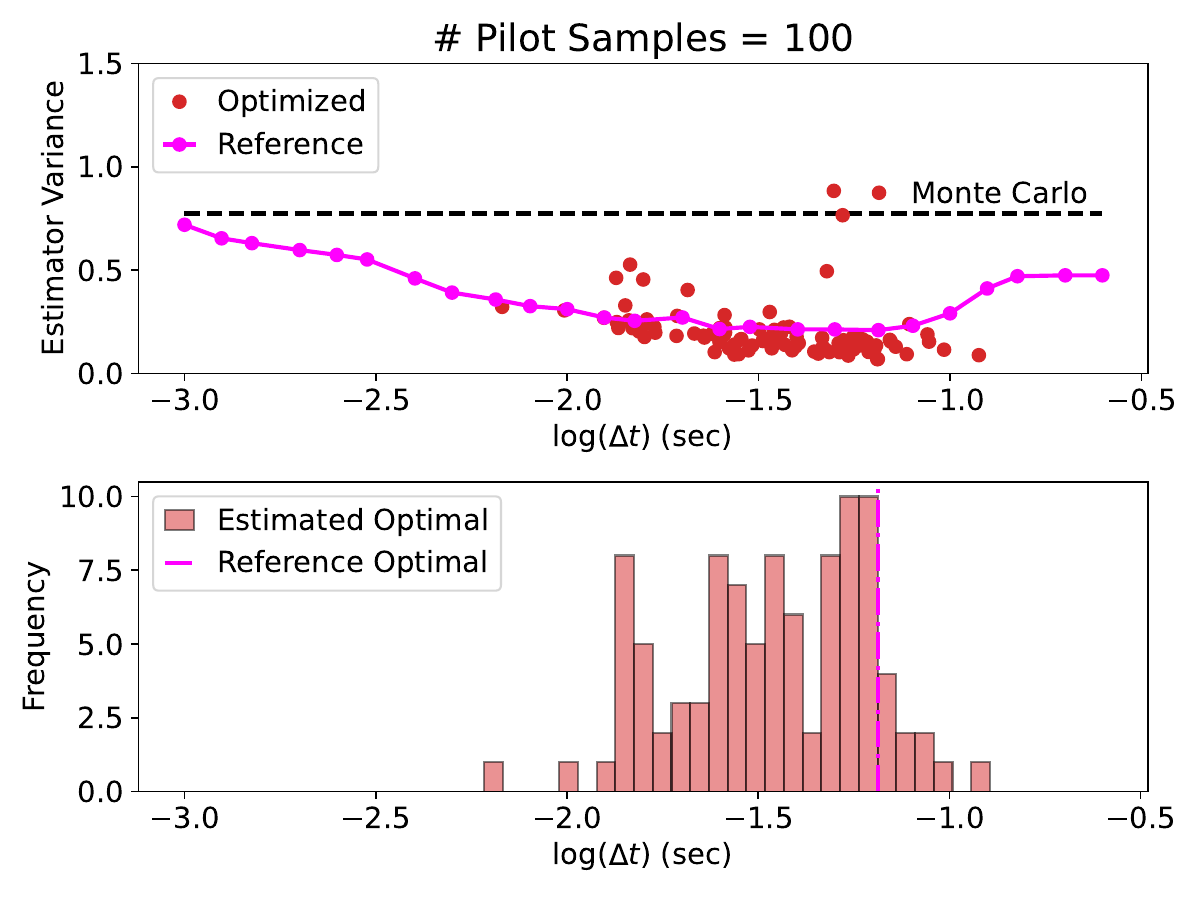}}
\caption{Model tuning optimization results for varying numbers of pilot samples with circular markers indicating the estimated optimal time step and estimator variance.}\label{fig:1d_opt_results_vs_num_pilot}
\end{figure}

\begin{figure} 
\centering 
\subfigure[]{\includegraphics[width=2.95in]{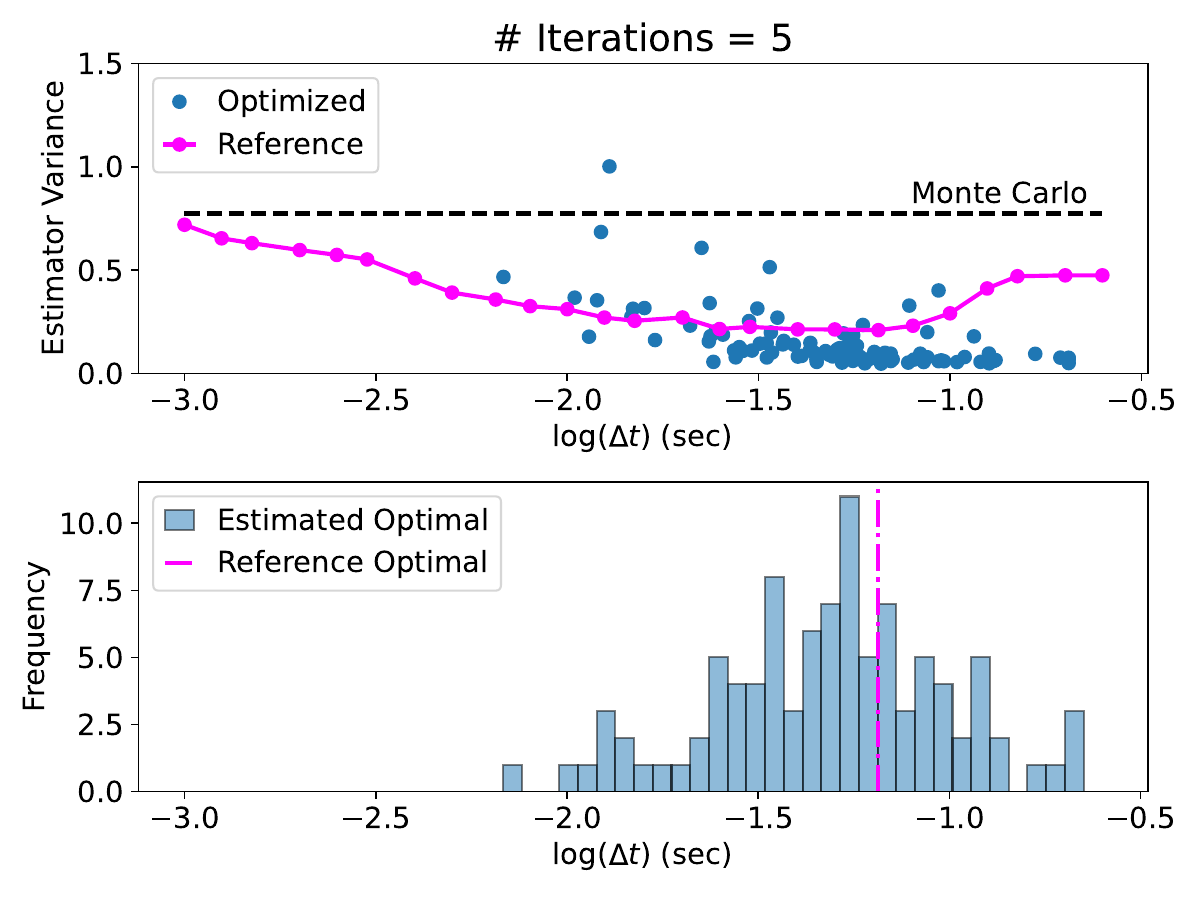}}
\subfigure[]{\includegraphics[width=2.95in]{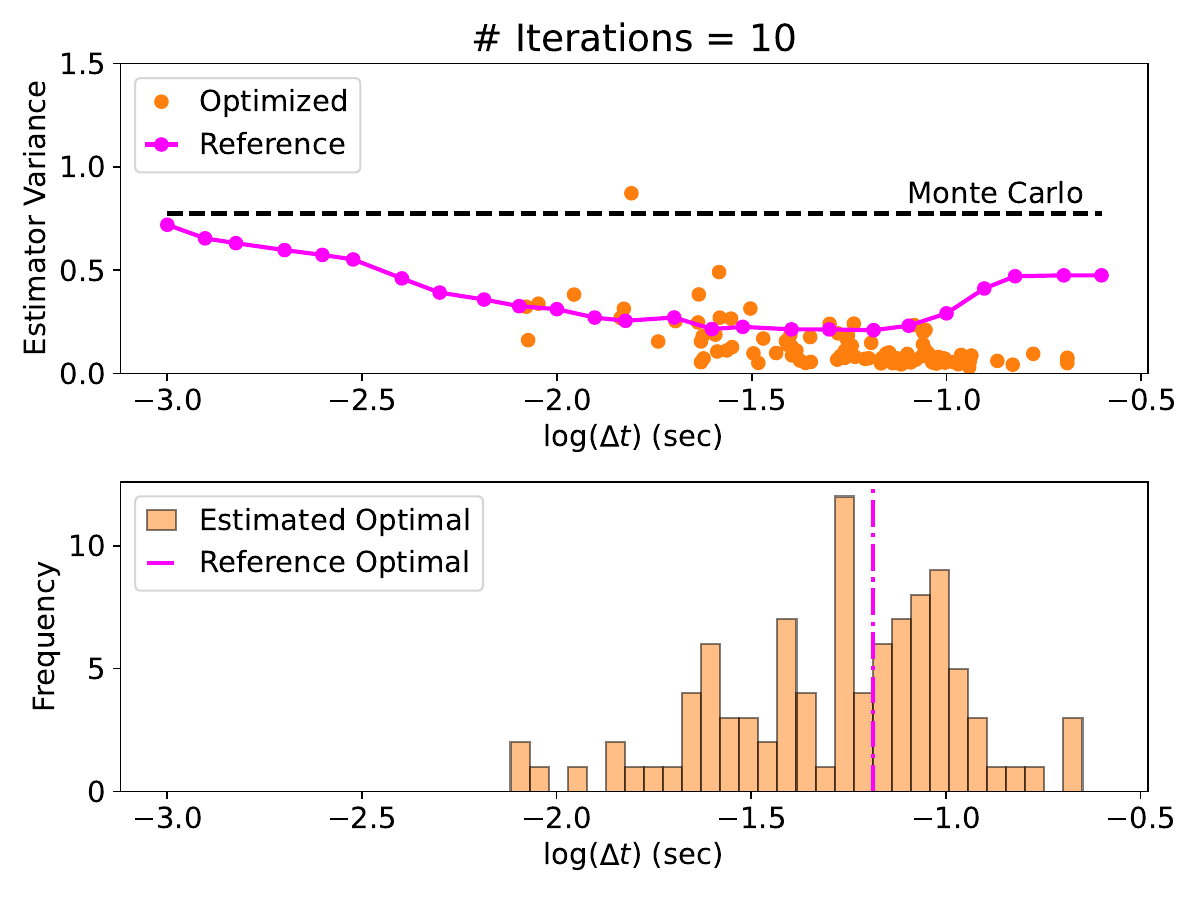}}
\subfigure[]{\includegraphics[width=2.95in]{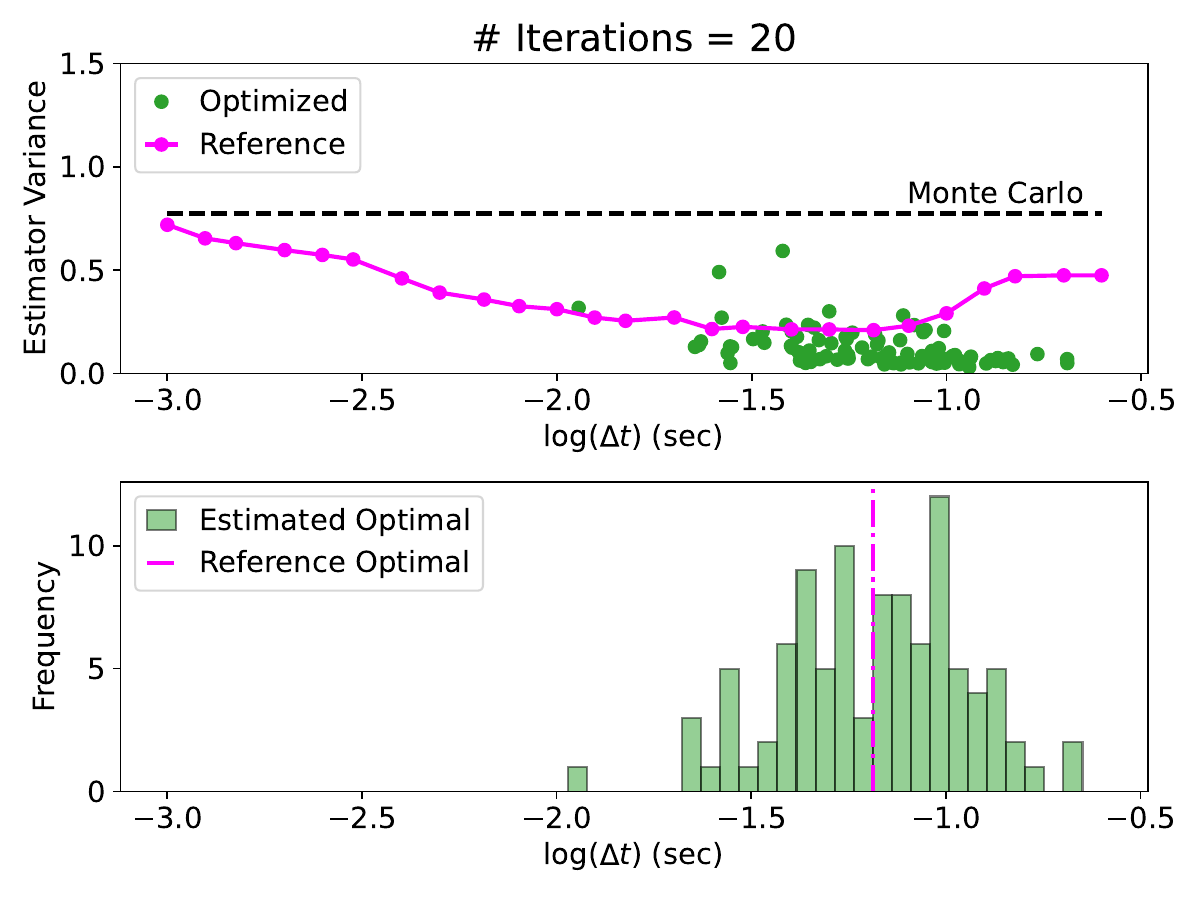}}
\caption{Model tuning optimization results for varying numbers of EGO iterations with circular markers indicating the estimated optimal time step and estimator variance.}\label{fig:1d_opt_results_vs_num_iteration}
\end{figure}

Detailed model tuning results for example 1 are shown in Figures \ref{fig:1d_opt_results_vs_num_pilot} and \ref{fig:1d_opt_results_vs_num_iteration} for varying numbers of pilot samples and EGO iterations, respectively. The top figures are scatter plots that show the estimated variance for each of the optimized time steps over 100 random trials. There is significant variability in the estimated variances, estimated with $N_{pilot}\leq100$, versus the reference variance curve that was estimated with $N_{pilot}=10k$. While Figure  \ref{fig:1d_opt_results_vs_num_pilot} shows the relative scatter in both the estimated variance and optimized time step decrease with increasing $N_{pilot}$, this variability is relatively unchanged with increasing $N_{iter}$ in Figure \ref{fig:1d_opt_results_vs_num_iteration} 

\section{Model Tuning Optimization Problem Noise Illustration}\label{sec:appendix_ex1_opt_noise}

\begin{figure} 
\centering 
\subfigure[]{\includegraphics[width=2.95in]{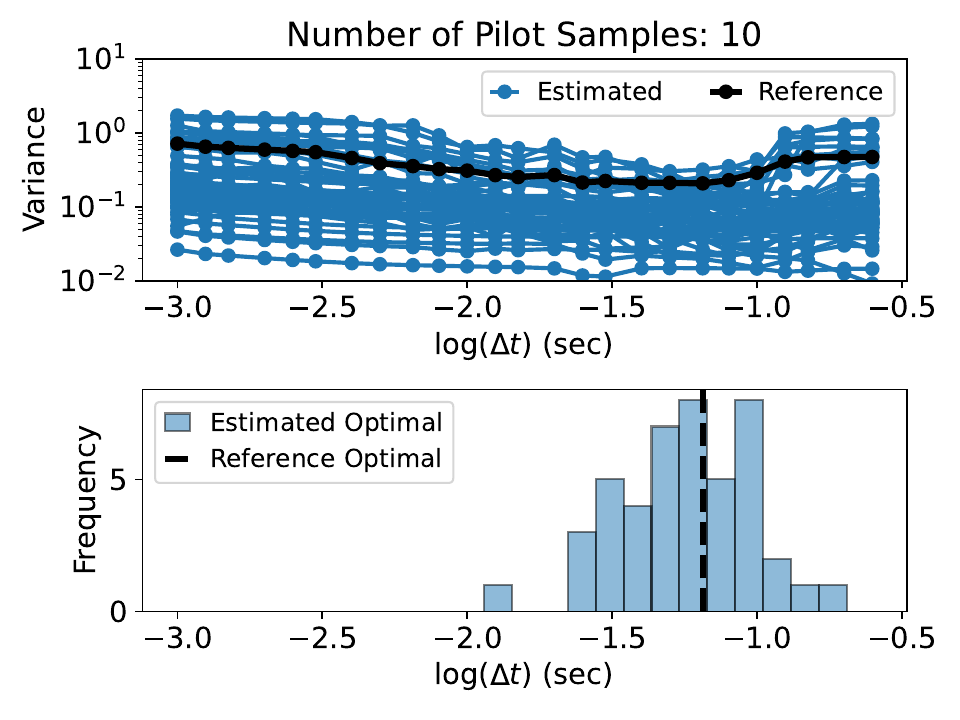}}
\subfigure[]{\includegraphics[width=2.95in]{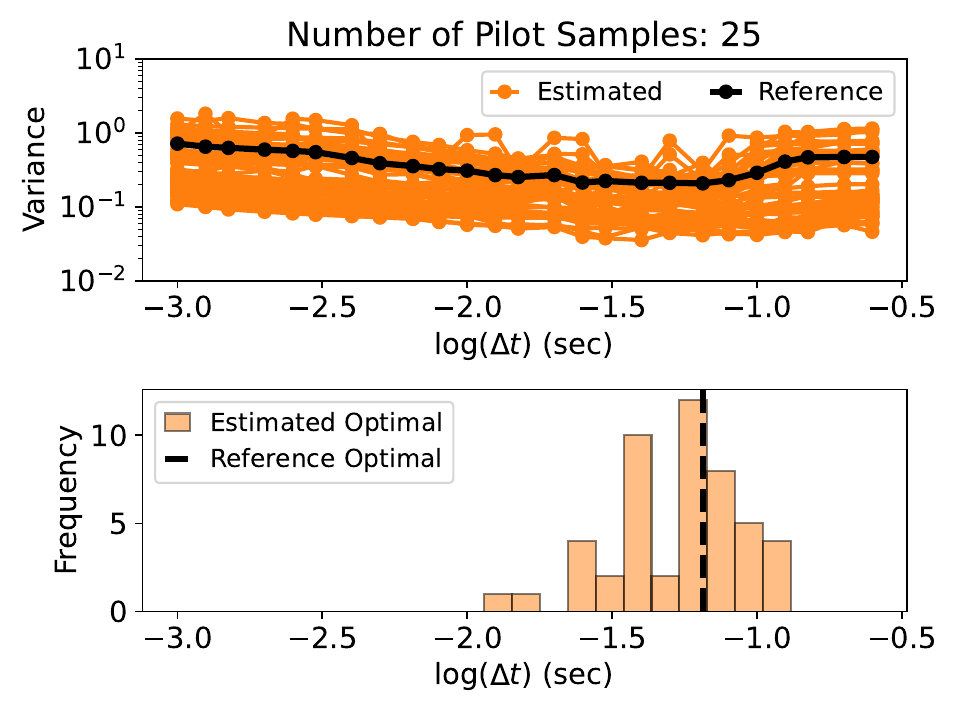}}
\subfigure[]{\includegraphics[width=2.95in]{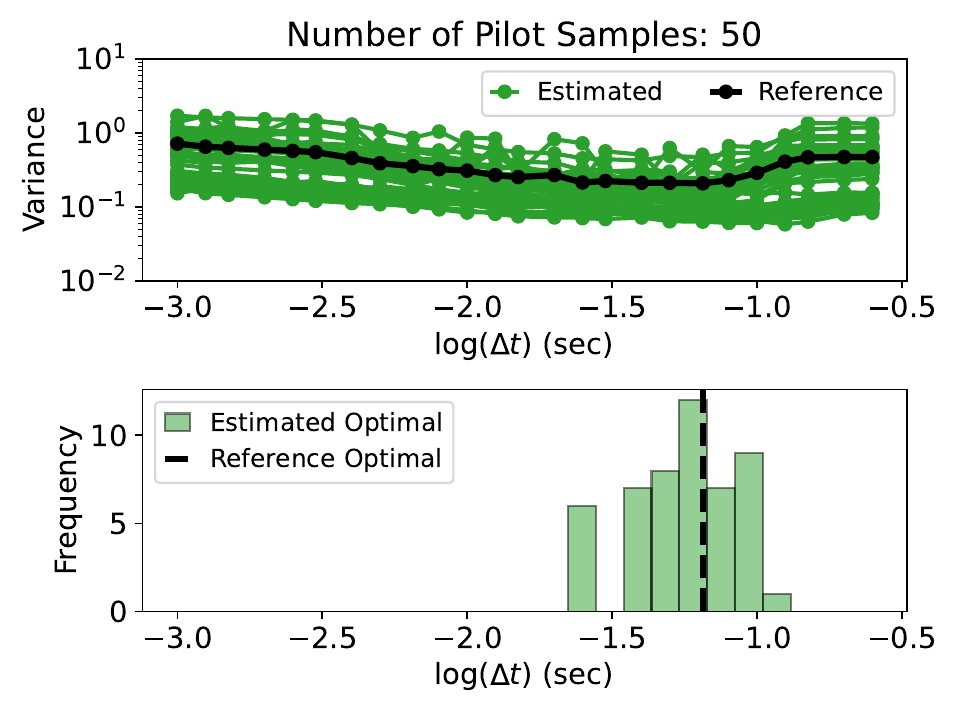}}
\subfigure[]{\includegraphics[width=2.95in]{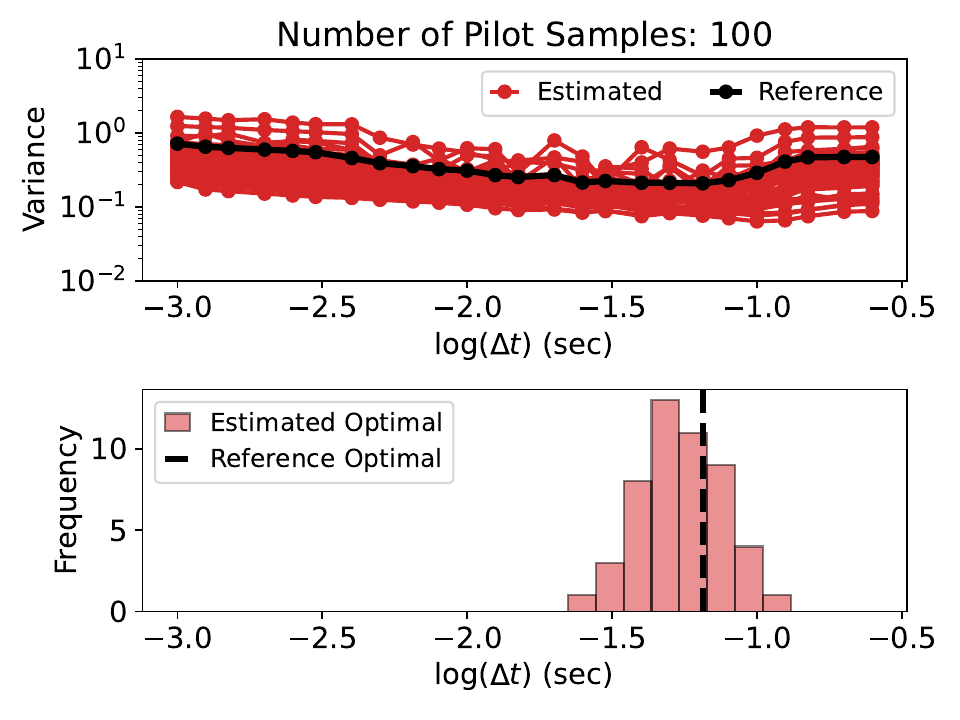}}
\caption{Variability in estimator variance versus $log(\Delta t)$ across 50 random trials when using a) 10, b) 25, c) 50, and d) 100 pilot samples.}\label{fig:1d_example_sweep_trials}
\end{figure}

This section helps illustrate the challenge posed by the noisy objective function for model tuning optimization for the ADEPT application. Figure \ref{fig:1d_example_sweep_trials} shows how estimates of the estimator variance (the objective function) vary across random trials for different numbers of pilot samples. Here, the estimator variance was evaluated on a grid of $25$ evenly spaced values of $log(\Delta t)$ for the time of flight QoI considered in Example 1. This evaluation was performed for $50$ random trials using $N_{pilot}=\{10, 25, 50, 100\}$ and compared to the reference estimator variance curve computed with $N_{pilot} = 10k$. Additionally, the optimal time step was found as the minimum across the $25$  $log(\Delta t)$ values considered for each trial. The figure shows significant variability in the vertical location of the estimator variance curves as well as the optimal time steps found (shown in the histograms in the lower plots). While the variability decreases as $N_{pilot}$ is increased, there is still a relatively wide range of optimal time step values for $N_{pilot}=100$. This illustrates that even for ideal optimizer performance (using a brute force sweep), it remains a challenge to optimize the estimator variance objective due to the noise associated with the ACV optimization problem and finite pilot sampling used.

\end{appendices}

\end{document}